\documentclass[a4paper,11pt]{article}
\usepackage{caption}
\usepackage{subcaption}
\usepackage{jheppub}
\usepackage{svg}
\usepackage{braket}
\usepackage{lineno}
\usepackage{jheppub}

\title{\boldmath Beta function without ultraviolet divergences}

\author[a,b,1]{Maxim Gritskov,}
\affiliation[a]{Krichever Center, Skolkovo Institute of Science and Technology, 121205, Moscow, Russia}
\affiliation[b]{Saint Petersburg State University,
Universitetskaya nab. 7/9, 199034 St. Petersburg, Russia}
\author[c]{Andrey Losev\,}
\affiliation[c]{Shanghai Institute for Mathematics and Interdisciplinary Sciences,\\
Building 3, 62 Weicheng Road, Yangpu District, 200433, Shanghai, China}

\emailAdd{maximgritskovvl@gmail.com}
\emailAdd{aslosev2@yandex.ru}

\abstract{In this paper, we construct the beta function in the functorial formulation of two-dimensional quantum field theories (FQFT). A key feature of this approach is the absence of ultraviolet divergences. We show that, nevertheless, in the FQFT perturbation theory, the local observables of deformed theories acquire logarithmic dimension, leading to a conformal anomaly. The beta function arises in the functorial approach as an infinitesimal transformation of the partition function under the variation of the metric's conformal factor, without ultraviolet divergences, UV cutoff, or the traditional renormalization procedure.}


\note{Corresponding author.}

\begin{document}
\maketitle
\flushbottom

\newpage
\section{Introduction}
\label{sec: Introduction}
There are three levels of understanding of quantum field theories and local observables in them. The first level is Feynman's formulation of QFT: the functional integral and its renormalization. The second level is the proto-algebraic approach. There, the first conceptual change occurs; namely, in this approach fields are not fundamental objects. Finally, the third level is the functorial quantum field theory. In this approach, we abandon the primary notion of a local observable at all and consider observables as derived objects.

In this section, we give a brief overview of each of these approaches to emphasize the key differences arising on the way from the first formulation to the third. Of course, there are many other attempts to answer the question of what quantum field theory is. Many of them are discussed in \cite{Snowmass}.
\subsection{First Level: Feynman's Approach to QFT}
\label{subsec: First Order}
The first level of understanding is the description of quantum field theory via the functional integral. In this setting, the key concept is the fundamental fields (and the linear structures on them). Fundamental fields are global sections $\Gamma(E,X)$ of a vector bundle $E\rightarrow X$ over spacetime $X$. We will demonstrate the Feynman approach with an example where the fundamental fields are just functions on spacetime $X$ (scalar field theory). We will denote the space of fields by $\mathrm{Func}(X)$.\footnote{We do not want to discuss in detail the integration domain in the functional integral approach, since this is a complicated issue even for scalar field theory. Therefore, we will use the universal notation $\mathrm{Func}(\cdot)$.} Furthermore, for simplicity, we assume that the spacetime manifold $X$ is just flat Euclidean space $\mathbb{R}^{n}$. Local observables are functionals $F[\phi(x),\partial\phi(x), \partial^{2}\phi(x),...]$ of the fundamental fields, polynomial in their derivatives. In the following, for brevity, we omit the dependence on higher derivatives of the fundamental fields $\phi$. The correlation function of local observables is given by
\begin{equation}
\label{eq:1.1}
\begin{aligned}
\braket{O_{1}(x_{1})\,...\,O_{n}(x_{n})}^{\mathrm{Feynman}}_{\mathrm{Naive}}&=\int_{\phi\,\in\,\mathrm{Func}(\mathbb{R}^{n})}\mathcal{D}\phi\,\mathrm{e}^{-S_{g}[\phi]}\left(F_{1}[\phi(x_{1})]\cdot ...\cdot F_{n}[\phi(x_{n})]\right)\,,
\end{aligned}
\end{equation}
where $S_{g}[\phi]$ is the classical action of the theory, depending on the fundamental fields $\phi$ and the set of coupling constants $g$. 

However, trying to interpret this formula literally leads to infinities (ultraviolet divergences). Therefore, it requires clarification, namely, the introduction of a renormalization procedure. In one version of the renormalization prescription, it means that the integration space and all parameters should be replaced by regularized ones:
\begin{equation}
\label{eq:1.2}
\begin{aligned}
\mathrm{Func}(\mathbb{R}^{n})\rightarrow\mathrm{Func}_{\Lambda}(\mathbb{R}^{n})=\left\{\phi(x)=\int_{p^{2}<\Lambda^{2}}\frac{\mathrm{d}^{n}p}{(2\pi)^{n}}\,\mathrm{e}^{\mathrm{i}p\cdot x}\cdot \hat{\phi}(p)\right\}\,,\\
g\rightarrow g(\Lambda)\,,
\qquad
F_{k}[\phi(x_{k}),\partial\phi(x_{k}),...]\rightarrow F_{k}^{(\Lambda)}[\phi(x_{k}),\partial\phi(x_{k}),...]\,.
\end{aligned}
\end{equation}
Then the dependence of all these objects on $\Lambda$ is chosen such that there exists the limit
\begin{equation}
\label{eq:1.3}
\begin{aligned}
\lim_{\Lambda\rightarrow \infty}\int_{\phi\,\in\,\mathrm{Func}_{\Lambda}(\mathbb{R}^{n})}\mathcal{D}\phi\,\mathrm{e}^{-S^{(\Lambda)}_{g(\Lambda)}[\phi]}\left(F^{(\Lambda)}_{1}[\phi(x_{1})]\cdot ...\cdot F^{(\Lambda)}_{n}[\phi(x_{n})]\right)\,.
\end{aligned}
\end{equation}
The value of this limit is called the renormalized correlator $\braket{...}_{\mathrm{Ren}.}^{\mathrm{Feynman}}$.

In conformal theories the renormalized correlation functions satisfy the following relations:
\begin{equation}
\label{eq:1.4}
\begin{aligned}
\partial_{\lambda}\braket{O_{a_{1}}(\lambda\cdot x_{1})\,...\,O_{a_{n}}(\lambda\cdot x_{n})}^{\mathrm{Feynman}}_{\mathrm{Ren}.}\big|_{\lambda=1}=\\=\sum_{k=1}^{n}\mathcal{D}^{b}_{a_{k}}\cdot\braket{O_{a_{1}}(x_{1})\,...\,O_{b}(x_{k})\,...\,O_{a_{n}}(x_{n})}^{\mathrm{Feynman}}_{\mathrm{Ren}.}\,.
\end{aligned}
\end{equation}
The linear operator $\mathcal{D}$ defined by equation~\eqref{eq:1.4} is called the dilatation generator.

Let us consider a simple example of a two-dimensional CFT, namely the free boson:
\begin{equation}
\label{eq:1.5}
\begin{aligned}
S[\phi]&=\int_{\mathbb{R}^{2}}\mathrm{d}\phi\wedge\star\,\mathrm{d}\phi\,.
\end{aligned}
\end{equation}
In this theory, we consider a local observable $O_{\alpha}$ given by $F_{\alpha}[\phi(x)]=\mathrm{e}^{\mathrm{i}\alpha \phi(x)}$. There is a well-known relation:
\begin{equation}
\label{eq:1.6}
\begin{aligned}
\braket{O_{\alpha}(x)O_{-\alpha}(y)}^{\mathrm{Feynman}}_{\mathrm{Reg}.}&=\int_{\phi\,\in\,\mathrm{Func}_{\Lambda}(\mathbb{R}^{2})}\mathcal{D}\phi\,\mathrm{e}^{-S[\phi]+\mathrm{i}\alpha\phi(x)-\mathrm{i}\alpha\phi(y)}\sim\frac{\mathrm{e}^{\alpha^{2}\log(\Lambda)}}{|x-y|^{\alpha^{2}}}\,.
\end{aligned}
\end{equation}
It immediately follows that $F_{\alpha}^{(\Lambda)}[\phi(x)]=\Lambda^{-\frac{\alpha^{2}}{2}}\,F_{\alpha}[\phi(x)]$. Then we obtain
\begin{equation}
\label{eq:1.7}
\begin{aligned}
\braket{O_{\alpha}(x)O_{-\alpha}(y)}^{\mathrm{Feynman}}_{\mathrm{Ren.}}&\sim\frac{1}{|x-y|^{\alpha^{2}}}\,.
\end{aligned}
\end{equation}
This correlator indeed satisfies the relation~\eqref{eq:1.4}:
\begin{equation}
\label{eq:1.8}
\begin{aligned}
\braket{O_{\alpha}(\lambda\cdot x)O_{-\alpha}(\lambda\cdot y)}^{\mathrm{Feynman}}_{\mathrm{Ren}.}&=\lambda^{-\alpha^{2}}\cdot \braket{O_{\alpha}(x)O_{-\alpha}(y)}^{\mathrm{Feynman}}_{\mathrm{Ren}.}\,.
\end{aligned}
\end{equation}

However, not all known quantum field theories have the Feynman formulation. For example, various lattice models in their critical points are quantum field theories without Lagrangian (functional integral) description.
\subsection{Second Level: Wightman’s Axiomatic Approach to QFT}
\label{subsec: Second Order}
It turns out that the notion of quantum field theory can be extended beyond the Feynman level. This extension is called Wightman’s quantum field theory \cite{AlgebraicQFT}, and in the case of two-dimensional conformal theories, it is realized through the BPZ axioms \cite{BPZ}. In this approach, we start with the vector space $\mathcal{O}$ of local observables $O_{a}$ with a distinguished observable called the stress-energy tensor $T$. Then we consider the configuration spaces $\hat{X}_{n}$, which are constructed from spacetime $X$ as $X^{n}\backslash\mathrm{Diagonal}(X^{n})$. By $\mathrm{Diagonal}(X^{n})$ we mean all such configurations of $n$ points in $X$ that at least two points collide. For simplicity, we will consider $\mathbb{R}^{2}$ as the spacetime $X$. The key difference from the Feynman approach is that we abandon the concept of a fundamental field. The fundamental objects are the correlation functions of the local observables, defined as linear maps:
\begin{equation}
\label{eq:1.9}
\begin{aligned}
\braket{\,}^{\mathrm{BPZ}}:\mathcal{O}^{\otimes n}&\rightarrow \mathrm{Func}(\hat{X}_{n})\,.
\end{aligned}
\end{equation}

In particular, the operator $\mathcal{D}$ is defined on the primary observables as follows:
\begin{equation}
\label{eq:1.10}
\begin{aligned}
\braket{T(x)\,O_{a}(y)}^{\mathrm{BPZ}}&=\frac{\mathcal{D}_{a}^{b}}{|x-y|^{2}}\cdot \braket{O_{b}(y)} ^{\mathrm{BPZ}}+\mathrm{less\,singular\,terms\, in}\,|x-y|\,.
\end{aligned}
\end{equation}

Just as in the Feynman's approach one can study deformation theory in the BPZ approach. Namely, one can deform the operation $\braket{\,}^{\mathrm{BPZ}}$ by set of local observables $\{O_{b}\}$ in the first order as follows
\begin{equation}
\label{eq:1.12}
\begin{aligned}
\braket{O_{a_{1}}(x_{1})\,...O_{a_{n}}(x_{n})}^{\mathrm{BPZ}}_{\mathrm{def}}=\braket{O_{a_{1}}(x_{1})\,...O_{a_{n}}(x_{n})}^{\mathrm{BPZ}}+\\ +\,g^{b}\int_{\mathbb{R}^{2}}\mathrm{d}y\,\braket{O_{a_{1}}(x_{1})\,...O_{a_{n}}(x_{n})\,O_{b}(y)}^{\mathrm{BPZ}}\,.
\end{aligned}
\end{equation}
Here we again encounter UV divergences arising from the singular behavior of the integrand. Once again, we need a regularization prescription. For example, consider $n=1$:
\begin{equation}
\label{eq:1.13}
\begin{aligned}
\braket{O_{a}(x)\,O_{b}(y)}^{\mathrm{BPZ}}&\sim\frac{C_{ab}^{\,c}}{|x-y|^{\Delta}}\cdot \braket{O_{c}(y)}^{\mathrm{BPZ}},\,x\rightarrow y\,.
\end{aligned}
\end{equation}
In this approach, the regularization prescription involves replacing the initial configuration space $(\mathbb{R}^{2})^{\times n}\backslash\mathrm{Diagonal}((\mathbb{R}^{2})^{\times n})$ with the regularized one:
\begin{equation}
\label{eq:1.14}
\begin{aligned}
\hat{X}_{n}^{(r)}&=(\mathbb{R}^{2})^{\times n}\backslash\mathrm{Tub}_{r}(\mathrm{Diagonal}((\mathbb{R}^{2})^{\times n}))\,,
\end{aligned}
\end{equation}
where $\mathrm{Tub}_{r}$ denotes the tubular neighborhood. In this case the integration~\eqref{eq:1.12} replaced by the integration over the plane $\mathbb{R}^{2}$ with the disk $D_{r}(x)$ cut out:
\begin{equation}
\label{eq:1.15}
\begin{aligned}
\int_{\mathbb{R}^{2}\backslash D_{r}(x)}\mathrm{d}y\,\braket{O_{a}(x)\,O_{b}(y)}^{\mathrm{BPZ}}&\sim r^{2-\Delta}\cdot C_{ab}^{\,c}\cdot \braket{O_{c}(y)}^{\mathrm{BPZ}}\,
\end{aligned}
\end{equation}
where $r^{0}$ should be understood as $\log(r)$. Then the UV divergence can be removed by redefining the space of local observables of the deformed theory, namely:
\begin{equation}
\label{eq:1.16}
\begin{aligned}
O_{a}&\rightarrow O_{a}^{(r)}=O_{a}-r^{2-\Delta}\cdot g^{b}\cdot C_{ab}^{\,c}\cdot O_{c}\,.
\end{aligned}
\end{equation}
For such a renormalized local observable, the operation $\braket{\,}_{\mathrm{def}}^{\mathrm{BPZ}}$ is well defined.

Note that in this approach, the deformation also requires a regularization prescription. When trying to construct a higher-order deformation, we will have to perform integration over a regularized configuration space~\eqref{eq:1.14}. However, integration will no longer be reduced to integration over a simple domain such as $\mathbb{R}^{2}\backslash D_{r}(x)$. In practice it might prove difficult to determine the correct region of integration \cite{LosevShifmanGamayun}.

\subsection{Third Level: Functorial QFT}
\label{subsec: Third Order}
Finally, we have reached the third level, namely the functorial QFT. This approach studies QFTs on compact spaces $X$ with boundary $\partial X=\Gamma$. This approach was first formulated by Segal in \cite{SegalFQFT}. To physicists the key idea of Segal's formulation is known as the cutting and gluing properties of partition functions. 

If the theory has a Feynman formulation, the partition function is defined as the functional on the boundary conditions space. Then for $\psi\in\mathrm{Func}(\Gamma)$ we have
\begin{equation}
\label{eq:1.17}
\begin{aligned}
\braket{\,}_{X}[\psi]&=\int_{\phi\,\in\,\mathrm{Func}(X):\,\phi|_{\Gamma}=\psi}\mathcal{D}\phi\,\mathrm{e}^{-S[\phi]}\,.
\end{aligned}
\end{equation}
In the Feynman approach, this functional has the well-known gluing property. Cut the spacetime $X$ along the curve $\gamma$. We obtain two manifolds $X_{1}$ and $X_{2}$ with the additional boundary components $\gamma$. Then there is the identity
\begin{equation}
\label{eq:1.18}
\begin{aligned}
\int_{\psi\,\in\,\mathrm{Func}(\gamma)}\mathcal{D}\psi\, \braket{\,}_{X_{1}}[\psi]\cdot \braket{\,}_{X_{2}}[\psi]=\braket{\,}_{X}\,.
\end{aligned}
\end{equation}

In the functorial approach, we abandon the functional integral and consider the gluing property of partition functions as an axiom. In $d=1$ (functorial quantum mechanics), this axiom is nothing but the Dirac evolution \cite{DiracEvolution}. This example is also discussed in appendix~\ref{C}.

In contrast to the BPZ axiomatics, in the functorial approach local observables are not fundamental but derived objects. In section~\ref{subsec: Local Observables in Functorial QFT} we will explain the notion of local observable in the functorial approach in detail.

The paper is organized as follows. In section~\ref{sec: Review of Functorial QFT} we formulate the axioms of functorial field theory. There, we define local observables in this approach and discuss their most important properties. In section~\ref{sec: Perturbation Theory in Functorial QFT} we construct perturbation theory and discuss a deep analogy between the renormalization procedure in quantum field theory and the transport of tangent vectors on a manifold. In section~\ref{sec: Beta Function in Two-Dimensional Theories} we study marginal deformations of two-dimensional conformal theories. We show that local observables in the deformed theory acquire a logarithmic dimension. Then we compute the anomalous dependence of the second-order partition function on the conformal factor of the metric.

\section{Review of Functorial QFT}
\label{sec: Review of Functorial QFT}
\subsection{Axioms of Functorial QFT}
\label{subsec: Axioms of Functorial QFT}
The main object in the functorial approach to QFT is the partition function on a manifold with boundary. In this paper, we study two-dimensional theories, but the functorial formulation holds for theories in spacetime of any dimension \cite{MultiDimensionalFQFT}. In appendix~\ref{C}, we consider an illustrative one-dimensional example, namely the functorial formulation of quantum mechanics. The partition function is defined axiomatically as follows: consider a Riemannian surface $X$ with metric tensor $h$ and a multicomponent boundary $\partial X=\Gamma_{1}\sqcup...\sqcup\Gamma_{n}$. Moreover, each component $\Gamma_{k}$ will have an additional top label ‘‘in’’ or ‘‘out’’. A surface with a fixed metric tensor $h$ is denoted by $X_{h}$. We associate a complex vector space $\mathcal{H}_{\Gamma_{k}^{(\mathrm{out})}}$ with each component $\Gamma_{k}^{(\mathrm{out})}$ (for $1\leq k\leq m$) and a dual space $\mathcal{H}_{\Gamma_{l}^{(\mathrm{in})}}^{*}$ with each component $\Gamma_{l}^{(\mathrm{in})}$ (for $m+1\leq l\leq n$). The partition function $\braket{\,}_{X}$ is an element of the tensor product 
\begin{equation}
\label{inout}
\begin{aligned}
\mathcal{H}_{\Gamma_{1}^{(\mathrm{out})}}\otimes...\otimes\,\mathcal{H}_{\Gamma_{m}^{(\mathrm{out})}} \otimes\mathcal{H}_{\Gamma_{m+1}^{(\mathrm{in})}}^{*}\otimes...\otimes\,\mathcal{H}_{\Gamma_{n}^{(\mathrm{in})}}^{*}\otimes\mathrm{Func}\left(\mathrm{Metrics}_{X}\right)
\end{aligned}
\end{equation}
that satisfies the product, cutting and equivariance axioms.

The product axiom states that if $X=X_{1}\sqcup X_{2}$, then $\braket{\,}_{X}=\braket{\,}_{X_{1}}\otimes \braket{\,}_{X_{2}}$. Let us formulate the cutting axiom. Let $\Gamma$ be a closed one-dimensional submanifold of $X_{h}$. Then we cut the surface $X_{h}$ along $\Gamma$ into two pieces: $X_{1,h_{1}}$ and $X_{2,h_2}$, where $h_{i}=h|_{X_{i}}$. We will denote the surface $X_{h}$ cut along $\Gamma$ into pieces $X_{1,h_{1}}$ and $X_{2,h_{2}}$ by $X^{\Gamma}_{h}=X_{1,h_{1}}\sqcup X_{2,h_{2}}$. The manifolds $X_{1,h_{1}}$ and $X_{2,h_{2}}$ have the common boundary component $\Gamma$, but for $X_{1,h_{1}}$ it is an in-boundary component and for $X_{2,h_{2}}$ it is an out-boundary component. The vector spaces associated with these boundary components are $\mathcal{H}_{\Gamma}^{\,*}$ (for $X_{1,h_{1}}$) and $\mathcal{H}_{\Gamma}$ (for $X_{2,h_{2}}$). There is a canonical pairing between them, which we will denote by $(\cdot)_{\Gamma}$. Then the cutting axiom requires that for any cutting contour $\Gamma$, the following relation holds:
\begin{equation}
\label{eq:2.1.1}
\begin{aligned}
\left(\braket{\,}_{X_{1,h_{1}}}\otimes \braket{\,}_{X_{2,h_{2}}}\right)_{\Gamma}&=\braket{\,}_{X_{h}}\,.
\end{aligned}
\end{equation}
Philosophically, one should look at equation~\eqref{eq:2.1.1} as an infinite system of quadratic equations that define the manifold of quantum field theories. In the one-dimensional case this system can be solved explicitly; see the appendix~\ref{C.1}. 

\begin{figure}[htbp]
\centering
\begin{subfigure}[t]{0.5\textwidth}
\centering
\includegraphics[width=.9\textwidth]{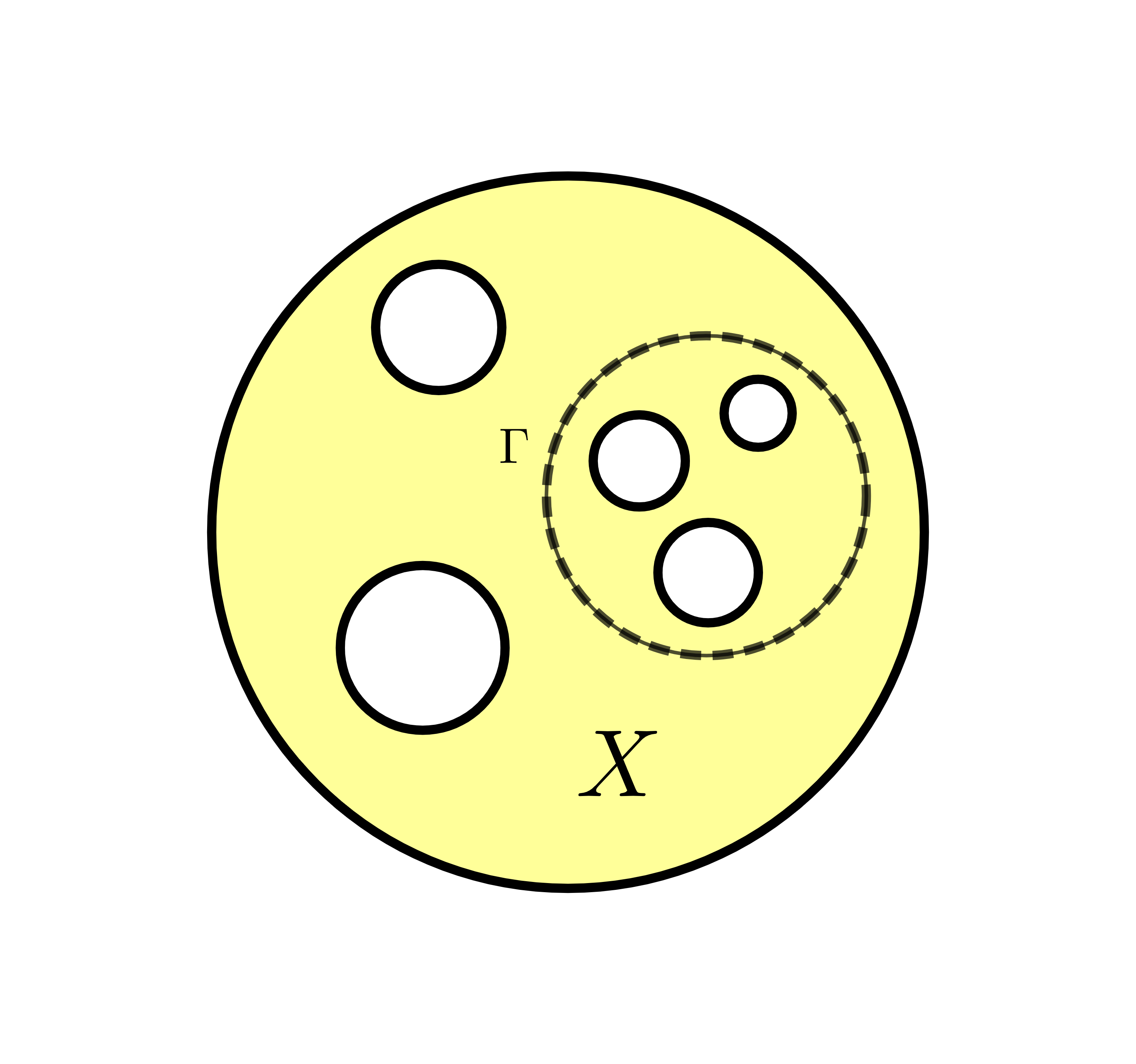}
\caption{}
\end{subfigure}%
\begin{subfigure}[t]{0.5\textwidth}
\centering
\includegraphics[width=.9\textwidth]{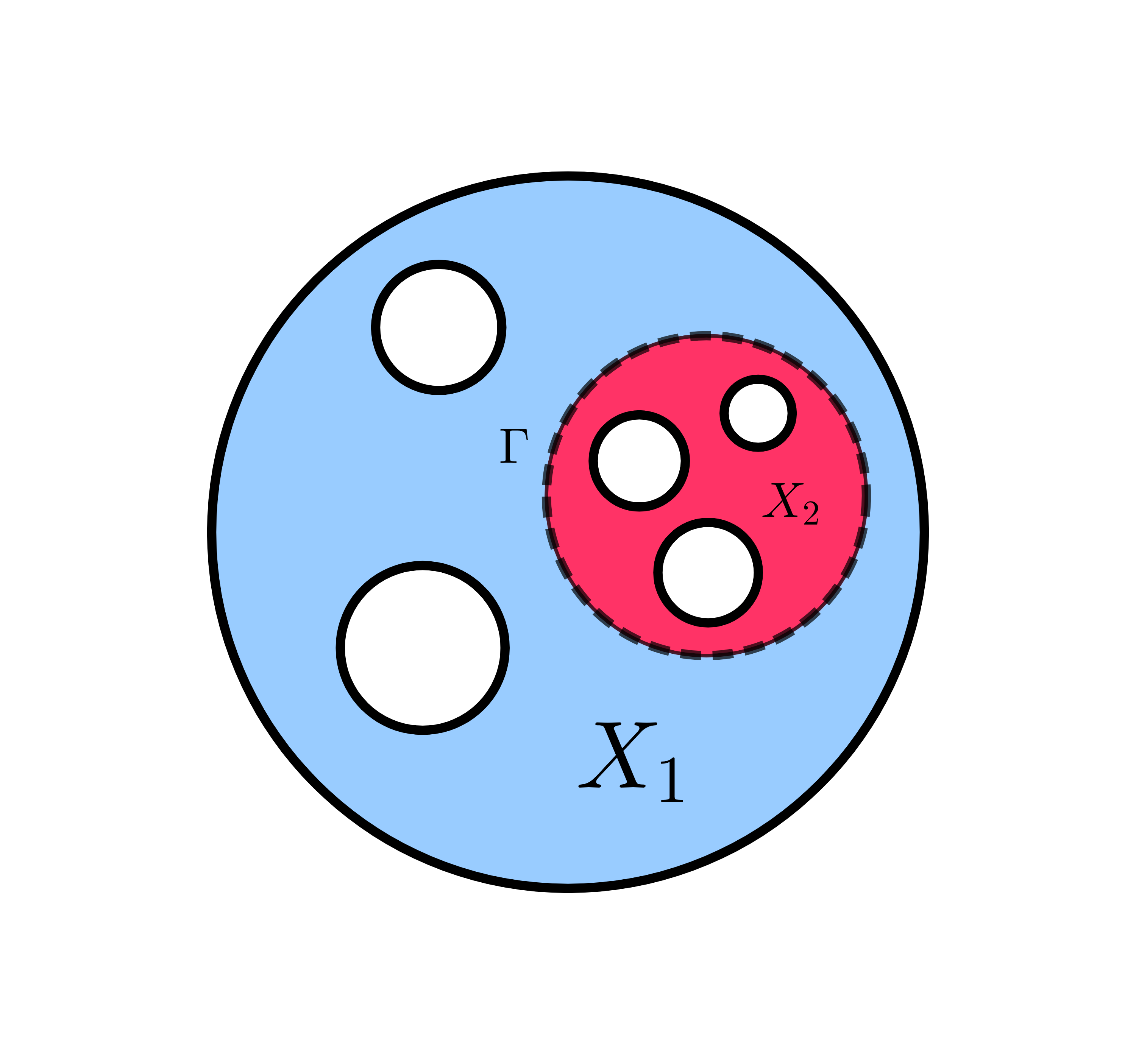}
\caption{}
\end{subfigure}
\caption{The yellow disk with holes $X$ shown in (a) is cut along the dotted line $\Gamma$, resulting in a union of the blue disk $X_{1}$ with a new hole and the red disk $X_{2}$, shown in (b).\label{fig:1}}
\end{figure}

Any compact twofold can be glued together from multiple disks with holes. Due to this fact, as well as local conformal symmetry and the cutting axiom, we can always assume that the spacetime manifold $X$ is a flat disk with a finite number of holes (see figure~\ref{fig:1}).

Considering figure~\ref{fig:1}, the cutting axiom states that the partition function on the yellow disk $X$ must coincide with the contraction of the partition functions on the blue disk $X_{1}$
and the red one $X_{2}$.

Now, let us formulate the equivariance axiom. Let $\mathrm{Diff}(X)$ be the group of diffeomorphisms of the manifold $X$ that preserve the boundary. Then we have a projective representation $\rho$ of $\mathrm{Diff}(X)|_{\partial X}$ on~\eqref{inout} such that for any $\phi\in\mathrm{Diff}(X)$ there is the identity 
\begin{equation}
\label{equivariance}
\begin{aligned}
\braket{\,}_{X_{\phi_{\star}h}} &=\rho(\phi|_{\partial X})\,\braket{\,}_{X_{h}}\,.
\end{aligned}
\end{equation}

In two dimensions, the metric is completely determined by the complex structure on the surface $X$ and the Weyl factor. All connected components of the boundaries in this case are isomorphic to $\mathrm{S}^{1}$, and the boundary condition spaces are projective representations of complexified $\mathrm{Diff}(\mathrm{S}^{1})$, or, which is the same, representations of the Virasoro algebra \cite{MnevLect}. Theories in which partition functions depend only on the complex structure but do not depend on the Weyl factor are called conformal \cite{ZZ}. Generally speaking, spaces associated with boundaries of different radii are not isomorphic to each other, but this is true for conformal theories. The projective representations account for the conformal anomaly arising from the nonzero central charge. In this paper, we will be interested in the phenomena of conformal anomaly in perturbed theories. Therefore, we will study conformal perturbation theory, that is, starting from conformal theory, we will perform perturbations by marginal observables and study the behavior of the deformed partition function and local observables under the action of Weil transformations.

Thus, in the constructed axiomatics, partition functions are considered as objects depending on the surface, its geometric data (metric), and boundary conditions. The axioms above were first formulated by Segal \cite{SegalFQFT}. From a mathematical point of view, these axioms mean that a partition function, as an object that takes as input a surface with a metric and provides as output an element of some vector space, realizes a functor from the category of geometrically enriched cobordisms to the category of vector spaces \cite{MnevLect}. That is why we call this formulation functorial quantum field theory (FQFT). If we discard
the geometric data, we obtain the well-known Atiyah axioms for topological QFT \cite{TQFT}.
\subsection{Local Observables in Functorial QFT}
\label{subsec: Local Observables in Functorial QFT}
In this section, we define local observables and their correlation functions. The key observation is that in FQFT they are not fundamental objects but arise as derived objects. Consider again the disk with holes $X$ from which we cut out a disk of radius $r$ centered at point $p\in X$, such that it is entirely contained in $X$.\footnote{On an arbitrary Riemannian surface $X_{h}$, this definition is naturally modified: we need to cut a disk of radius $r$ in the sense of the metric $h$.} We will denote such a disk by $D_{r}\left(p\right)$. The surface $X\backslash D_{r}\left(p\right)$ has an additional boundary component $\Gamma=\partial D_{r}\left(p\right)$. Consider the family of vectors $v_{r}\in\mathcal{H}_{\partial D_{r}\left(p\right)}$, depending on $r$. We will call a family $v_{r}$ good if the following limit exists:
\begin{equation}
\label{eq:2.2.2}
\begin{aligned}
\lim_{r\to 0}\left\{\left(\braket{\,}_{X\backslash D_{r}\left(p\right)}\otimes v_{r}\right)_{\partial D_{r}\left(p\right)}\right\}\,.
\end{aligned}
\end{equation}
We will call a good family $v_{r}$ a null family if the limit~\eqref{eq:2.2.2} is zero.

It is clear that because of the linearity of this definition in the family $v_{r}$, the right-hand side of the equation does not change if we add a null family to $v_{r}$. We will consider two good families to be equivalent if they differ by adding a null family. Therefore, the right-hand side of~\eqref{eq:2.2.2} depends only on $O_{v}=\left[v_{r}\right]_{\sim}$ - an equivalence class of the family $v_{r}$. The space of local observables $\{O_{v}\}$ at point $p$ is defined as the quotient space of good families over null families. 

For good families, we define the correlator $\braket{O_{v}(p)}_{X}$ of the local observable $O_{v}$ at the point $p\in X$ as the value of the limit~\eqref{eq:2.2.2}. Since local observables are implemented by the representatives of good families equivalence classes, we will often identify good families $v_{r}$ with local observables $O_{v}$.

Local observables depend on the theory, and we will see in the following sections that the key question is how the space of observables transforms under small deformations of the theory.

It is clear that we can sequentially put several local observables on the surface $X$. Thus, we define the correlation function of several local observables $\braket{O_{1}(p_{1})\,O_{2}(p_{2})\,...\,O_{n}(p_{n})}_{X}$ as the result of sequentially cutting disks at points $p_{1},p_{2},...,p_{n}$ and inserting there good families $v_{1,r},v_{2,r},...,v_{n,r}$ corresponding to local observables $O_{1},O_{2},...,O_{n}$, respectively. In the next section, we will study the behavior of such correlators when two observables collide.

Now, we formulate some important properties of local observables in FQFT.
\begin{itemize}
\item Let $X$ be cut along $\Gamma$ such that $X^{\Gamma}=X_{1}\sqcup X_{2}$ and $v_{r}$ is a good family for the partition function $\braket{\,}_{X_{2}\backslash D_{r}(p)}$. Then $v_{r}$ is a good family also for $\braket{\,}_{X\backslash D_{r}(p)}$, and
\begin{equation}
\label{Property 1.1}
\begin{aligned}
\braket{O_{v}(p)}_{X}&=\left(\braket{\,}_{X_{1}}\otimes\braket{O_{v}(p)}_{X_{2}}\right)_{\Gamma}\,.
\end{aligned}
\end{equation}
In fact, for the surface $X$ cut along $\Gamma$, we cut a disk of a small radius $r$ around the point $p$, so that it is contained entirely in the component $X_{2}$. Then the identity holds:
\begin{equation}
\label{Propery 1.2}
\begin{aligned}
\braket{\,}_{X\backslash D_{r}(p)}&=\left(\braket{\,}_{X_{1}}\otimes\braket{\,}_{X_{2}\backslash D_{r}(p)}\right)_{\Gamma}\,.
\end{aligned}
\end{equation}
If there exists the limit
\begin{equation}
\label{Property 1.3}
\begin{aligned}
\braket{O_{v}(p)}_{X_{2}}&=\lim_{r\to 0}\left\{\left(\braket{\,}_{X_{2}\backslash D_{r}(p)}\otimes v_{r}\right)_{\partial D_{r}(p)}\right\}\,,
\end{aligned}
\end{equation}
then there exists also the limit
\begin{equation}
\label{Property 1.4}
\begin{aligned}
\braket{O_{v}(p)}_{X}&=\lim_{r\to 0}\left\{\left(\braket{\,}_{X\backslash D_{r}(p)}\otimes v_{r}\right)_{\partial D_{r}(p)}\right\}\,,
\end{aligned}
\end{equation}
and the identity~\eqref{Property 1.1} holds. We will often use this property in the cases where $X_{2}$ is a disk centered at the point $p$, i.e., $X_{1}=X\backslash D_{R}(p),\,X_{2}=D_{R}(p)$ and $\Gamma=\partial D_{R}(p)$.
\item Consider the correlation function $\braket{O_{v}(0)}_{D_{r}}$ of a local observable $O_{v}$ inserted into the center $p=0$ of the disk $D_{r}$ and another partition function $\braket{\,}_{D_{R}}$ for $R>r$. Then, the correlation function $\braket{O_{v}(0)}_{D_{r}}\in\mathcal{H}_{\partial D_{r}}$ is a good family for partition function $\braket{\,}_{D_{R}\backslash D_{r}}$ on the annulus $D_{R}\backslash D_{r}$ and $\braket{O_{v}(0)}_{D_{{r}}}$ is equivalent to $v_{r}$ as a good family:
\begin{equation}
\label{Property 2.1}
\begin{aligned}
\lim_{r\to 0}\left\{\left(\braket{\,}_{D_{R}\backslash D_{r}}\otimes\braket{O_{v}(0)}_{D_{r}}\right)_{\partial D_{r}}\right\}&=\lim_{r\to 0}\left\{\left(\braket{\,}_{D_{R}\backslash D_{r}}\otimes v_{r}\right)_{\partial D_{r}}\right\}\,.
\end{aligned}
\end{equation}
In fact, according to the previous property, there is the identity
\begin{equation}
\label{Property 2.2}
\begin{aligned}
\braket{O_{v}(0)}_{D_{R}}&=\left(\braket{\,}_{D_{R}\backslash D_{r}}\otimes\braket{O_{v}(0)}_{D_{{r}}}\right)_{\partial D_{r}}\,.
\end{aligned}
\end{equation}
Therefore, the limit on the left side of equation~\eqref{Property 2.1} exists. The limit on the right-hand side of~\eqref{Property 2.1} exists under the assumption that $v_{r}$ is a good family corresponding to the observable $O_{v}$. By definition, it is equal to $\braket{O_{v}(0)}_{D_{R}}$.
\item The space of local observables in conformal theory has an important family of automorphisms related to scale transformations. Consider $\lambda>1$ and the partition function $\braket{\,}_{D_{\lambda\cdot r}\backslash D_{r}}$. We define the family of operators $\mathrm{Dil}_{\lambda}: \mathcal{H}_{\partial D_{r}}\rightarrow \mathcal{H}_{\partial D_{\lambda\cdot r}}$, parameterized by $\lambda$, as follows. Let $v\in\mathcal{H}_{\partial D_{r}}$, then
\begin{equation}
\label{Property 3.1}
\begin{aligned}
\mathrm{Dil}_{\lambda}v&=\left(\braket{\,}_{D_{\lambda\cdot r}\backslash D_{r}}\otimes v\right)_{\partial D_{r}}\in\mathcal{H}_{\partial D_{\lambda\cdot r}}\,.
\end{aligned}
\end{equation}
Since the theory is conformal, there is a canonical isomorphism $\mathcal{H}_{\partial D_{\lambda\cdot r}}\simeq\mathcal{H}_{\partial D_{r}}$. So, the operator $\mathrm{Dil}_{\lambda}$ is just an endomorphism of the space $\mathcal{H}_{\partial D_{r}}$.

Now, if the theory has conformal symmetry, then the partition function $\braket{\,}_{D_{\lambda\cdot r}\backslash D_{r}}$ does not depend on $r$. This is because $D_{R}\backslash D_{r}$ is conformally equivalent to a cylinder with length $\mathrm{log}(R/r)$, which is the only geometric datum for the cylinder. In this case, $\mathrm{Dil}_{\lambda}$ does not depend on $r$ and defines a $\lambda$-parametrized family of automorphisms of the space of local observables.\footnote{This means that it maps good families into other good families, and null families into null families.}
\end{itemize}

We will call the local observable $O_{v}$ conformal with scaling dimension $\Delta$ if
\begin{equation}
\label{Property 3.3}
\begin{aligned}
\mathrm{Dil}_{\lambda}\braket{O_{v}(0)}_{D_{r}}&=\lambda^{-\Delta}\cdot\braket{O_{v}(0)}_{D_{r}}\,.
\end{aligned}
\end{equation}
Usually in the space of local observables of a CFT there exists a basis of conformal observables. This means that the dilatation generator $\mathcal{D}=-\partial_{\lambda}\mathrm{Dil}_{\lambda}|_{\lambda=1}$ is a diagonalizable operator.\footnote{The dimensions of local observables are eigenvalues of the operator $\mathcal{D}$.} However, there are known examples of so-called logarithmic conformal field theories \cite{LosevFrenkelNekrasov2}, in which the dilatation generator $\mathcal{D}$ has a Jordan normal form.

It can be shown that in two-dimensional conformal theories the generator $\mathcal{D}$ factorizes into holomorphic and antiholomorphic terms. So, there are two dimensions defined: holomorphic $h$ and antiholomorphic $\bar{h}$, so that $\Delta=h+\bar{h}$. The difference $S=h-\bar{h}$ is called spin.

For our further purposes, we will need the notion of primary observables. We will call a conformal observable $O_{v}$ a primary observable if its correlation function $\braket{O_{v}(p)}_{X}$ on the surface $X$ without boundary transforms as tensor density under the action of diffeomorphisms.
\subsection{Operator Product Expansion}
\label{subsec: Operator Product Expansion}
In this section, we discuss the idea of how a new observable can be obtained from two known local observables. Consider the correlation function $\braket{O_{a}(q)O_{b}(0)}_{D_{r}}$ of local observables $v_{a,r'}$ and $v_{b,r'}$ on a disk $D_{r}\subset D_{R}$, where $\sqrt{q\cdot\bar{q}}=r/2$. In the previous section, we observed that if the local observable $O_{a}$ was absent, the resulting correlator would be a good family for the partition function $\braket{\,}_{D_{R}}$. However, with the addition of another local observable, the family may no longer be good:
\begin{equation}
\begin{aligned}
\label{eq:2.3.1}
\left(\braket{\,}_{D_{R}\backslash D_{r}}\otimes\braket{O_{a}\left(q\right)O_{b}\left(0\right)}_{D_{r}}\right)_{\partial D_{r}}&\sim q^{-\Delta_{1}^{a,b}}\cdot (\bar{q})^{-\bar{\Delta}_{1}^{a,b}}\,.
\end{aligned}
\end{equation}

Nevertheless, it still defines a new local observable: as a good family, let us choose the vector $q^{\Delta_{1}^{a,b}}\cdot(\bar{q})^{\bar{\Delta}_{1}^{a,b}}\cdot \braket{O_{a}\left(q\right)O_{b}\left(0\right)}_{D_{r}}$. This family defines a local observable, which must be a linear combination of already known local observables, i.e.
\begin{equation}
\begin{aligned}
\label{eq:2.3.2}
q^{\Delta_{1}^{a,b}}\cdot (\bar{q})^{\bar{\Delta}_{1}^{a,b}}\cdot \braket{O_{a}\left(q\right)O_{b}\left(0\right)}_{D_{r}}&\sim C_{a,b}^{1,c}\braket{O_{c}\left(0\right)}_{D_{r}}\,.
\end{aligned}
\end{equation}
We can now consider the following sequence as the candidate for a good family:
\begin{equation}
\label{eq:2.3.3}
\begin{aligned}
\tilde{v}_{r}&=\braket{O_{a}\left(q\right)O_{b}\left(0\right)}_{D_{r}}-q^{-\Delta_{1}^{a,b}}\cdot (\bar{q})^{-\bar{\Delta}_{1}^{a,b}}\cdot C_{a,b}^{1,c}\braket{O_{c}\left(0\right)}_{D_{r}}\,.
\end{aligned}
\end{equation}
Now, notice that
\begin{equation}
\label{eq:2.3.3.1}
\begin{aligned}
\left(\braket{\,}_{D_{R}\backslash D_{r}}\otimes\tilde{v}_{r}\right)_{\partial D_{r}}&\sim q^{-\Delta_{2}^{a,b}}\cdot (\bar{q})^{-\bar{\Delta}_{2}^{a,b}}\,,
\end{aligned}
\end{equation}
where $\Delta_{2}^{a,b}<\Delta_{1}^{a,b}$. Thus, we obtain a family leading to a new local observable:
\begin{equation}
\label{eq:2.3.4}
\begin{aligned}
q^{\Delta_{2}^{a,b}}\cdot (\bar{q})^{\bar{\Delta}_{2}^{a,b}}\cdot\tilde{v}_{r}&\sim C_{a,b}^{2,c}\braket{O_{c}\left(0\right)}_{D_{r}}\,.
\end{aligned}
\end{equation}
We will repeat this procedure multiple times until finally, after successive subtractions, we obtain a good family. Thus, we obtain the following decomposition of the two-point correlator into one-point correlators:
\begin{equation}
\label{eq:2.3.5}
\begin{aligned}
\braket{O_{a}\left(q\right)O_{b}\left(0\right)}_{D_{r}}=q^{-\Delta_{1}^{a,b}}\cdot(\bar{q})^{-\bar{\Delta}_{1}^{a,b}}\cdot C_{a,b}^{1,c}\braket{O_{c}\left(0\right)}_{D_{r}}+\\+\,q^{-\Delta_{2}^{a,b}}\cdot(\bar{q})^{-\bar{\Delta}_{2}^{a,b}}\cdot C_{a,b}^{2,c}\braket{O_{c}\left(0\right)}_{D_{r}}+\,...
\end{aligned}
\end{equation}
Historically, this decomposition was called the operator product expansion (OPE). The local conformal symmetry allows us to refine the formulas for the OPE; see appendix~\ref{A.1}. An example of OPE construction in the functorial framework is discussed in the appendix~\ref{B.2}. 

The following is worth noting. For any theory on the surface $X$ with metric $h$ we can construct the observable $T$ whose correlator will be given by the formula
\begin{equation}
\label{Stress-Energy}
\begin{aligned}
\braket{T(p)}_{X_{h}}&=\delta_{h(p)}\braket{\,}_{X_{h}}\,,
\end{aligned}
\end{equation}
where $\delta_{h(p)}$ is the variational derivative along the metric tensor at point $p$. This local observable is called the stress-energy tensor. The space of local observables of a conformal FQFT together with the stress-energy tensor~\eqref{Stress-Energy} satisfy the Belavin-Polyakov-Zamolodchikov axioms \cite{BPZ}.

Further, we will see that it is the presence of singular terms in the operator product expansion that leads to a non-trivial change in the space of local observables under deformation, and the OPE coefficients determine the renormalization group data.
\section{Perturbation Theory in Functorial QFT}
\label{sec: Perturbation Theory in Functorial QFT}
\subsection{Deformation of the Partition Function}
\label{subsec: Deformation of the Partition Function}
In the previous section, we have formulated the axioms defining QFT on a Riemannian manifold $X$ as a theory of partition functions $\braket{\,}_{X}$ satisfying the FQFT axioms. However, it is now necessary to develop perturbation theory within the framework of these axioms. 

Perturbation theory in the functorial approach is constructed using deformations. This construction has a clear geometric interpretation, which will be discussed in the next section. We expect that the deformational approach to perturbation theory has a serious computational advantage, since it allows us to solve some recurrence relations that are not clear in the traditional approach. For details, see \cite{LosevGritskov}.

We define the deformation of the partition function $\braket{\,}_{X}$ using the local observable $v_{r}$ by the formula:
\begin{equation}
\label{eq:3.1.1}
\begin{aligned}
\braket{\,}_{X}^{\left(\mathrm{def}\right)}\left(g,O_{v}\right)&=\braket{\,}_{X}+g\cdot\int_{p\,\in\,X}\mathrm{d}\mu_{p}\,\braket{O_{v}\left(p\right)}_{X}\,.
\end{aligned}
\end{equation}
Here, the measure $\mathrm{d}\mu_{p}$ is chosen in such a way that the deformation is invariant with respect to diffeomorphisms. We give an expression for this measure in the case of a deformation of the conformal theory by a primary spinless observable with conformal dimension $\Delta$. Choosing a holomorphic chart in which the coordinates of the point $p$ are $(z,\bar{z})$ and metric tensor takes the form $h=h_{z\bar{z}}\cdot(\mathrm{d}z\otimes\mathrm{d}\bar{z}+\mathrm{d}\bar{z}\otimes\mathrm{d}z)$, this measure is given by
\begin{equation}
\label{eq:3.1.2}
\begin{aligned}
\mathrm{d}\mu_{p}&=\frac{\mathrm{d}\bar{z}\wedge\mathrm{d}z}{4\pi \mathrm{i}}\cdot h_{z\bar{z}}^{\left(\frac{2-\Delta}{2}\right)}\,.
\end{aligned}
\end{equation}

It is easy to see that formula $\left(\ref{eq:3.1.1}\right)$ really defines a partition function, that is, it satisfies the cutting axiom~\eqref{eq:2.1.1} up to the terms of order $g^{2}$. As usual, we cut $X$ along $\Gamma$, so the initial surface is the union of two components: $X=X_{1}\sqcup X_{2}$. Then we obtain
\begin{equation}
\label{eq:3.1.3}
\begin{aligned}
\left(\braket{\,}_{X_{1}}^{\left(\mathrm{def}\right)}\left(g,O_{v}\right)\otimes \braket{\,}_{X_{2}}^{\left(\mathrm{def}\right)}\left(g,O_{v}\right)\right)_{\Gamma} = \left(\braket{\,}_{X_{1}}\otimes\braket{\,}_{X_{2}}\right)_{\Gamma}+
\\
+\,g\cdot\int_{p\,\in \,X_{1}}\mathrm{d}\mu_{p}\left(\braket{O_{v}(p)}_{X_{1}}\otimes \braket{\,}_{X_{2}}\right)_{\Gamma} + g\cdot\int_{p\,\in\,X_{2}} \mathrm{d}\mu_{p} \left(\braket{\,}_{X_{1}}\otimes \braket{O_{v}(p)}_{X_{2}}\right)_{\Gamma} =
\\
 =\braket{\,}_{X}+ g \cdot\int_{p\,\in\,X_{1}} \mathrm{d}\mu_{p} \, \braket{O_{v}(p)}_{X} + g \cdot\int_{p\,\in \,X_{2}} \mathrm{d}\mu_{p} \,\braket{O_{v}(p)}_{X} = \braket{\,}_{X}^{\left(\mathrm{def}\right)}\left(g,O_{v}\right)\,.
\end{aligned}
\end{equation}
Thus, the cutting axiom is satisfied. It is clear that one can equally well consider deformations of the initial theory by a linear combination of several local observables with different coupling constants.

We will call the deformations by primary spinless local observables of dimension $\Delta=2$ marginal, and the corresponding local observables will be called primary marginal. A well-known example of marginal deformations are current-current type deformations \cite{LosevShifmanGamayun, LosevShifmanGamayun2, Curcur}. At first sight, they do not break conformal symmetry; i.e., the deformed partition function also corresponds to some conformal field theory. This is due to the fact that the metric enters the deformed theory with degree $2-\Delta=0$. Then, for marginal deformations, the formula $(\ref{eq:3.1.1})$ for the QFT partition function on the Riemannian surface does not depend on the Weyl factor.

Clearly, the double deformation also satisfies the FQFT axioms by construction. Thus, the higher-order perturbation theory is constructed as a multiple deformation of the initial theory by local observables. However, we need to be careful at this point.

\subsection{Local Observables in the Deformed Theory}
\label{subsec: Local Observables of Deformed Theory}
The point is that, in general, the space of local observables would change under deformation of the theory. This circumstance is connected with the fact that the set of all quantum field theories is defined by the system of non-linear equations~\eqref{eq:2.1.1}, while the space of local observables is defined by the linear condition~\eqref{eq:2.2.2}. In fact, consider a deformed theory with partition function $\braket{\,}_{X}^{\left(\mathrm{def}\right)}(g,O_{a})$ and cut a disk of radius $r$ centered at the point $p$. Let us try to insert there the family $v_{b}$ from the observables space of the unperturbed theory:
\begin{equation}
\label{eq:3.2.1}
\begin{aligned}
\left(\braket{\,}^{\left(\mathrm{def}\right)}_{X\backslash D_{r}\left(p\right)}(g, O_{a})\otimes v_{b,r}\right)_{\partial D_{r}\left(p\right)}&=\braket{O_{b}(p)}_{X}\,+g\cdot\int_{q\,\in\,X\backslash D_{r}(p)}\mathrm{d}\mu_{q}\braket{O_{a}(q)O_{b}(p)}_{X}\,.
\end{aligned}
\end{equation}
However, it was shown in the section~\ref{subsec: Local Observables in Functorial QFT} that the integrand here has a singular behavior:
\begin{equation}
\label{eq:3.2.2}
\begin{aligned}
\braket{O_{a}(q)O_{b}(p)}_{X}\sim r_{pq}^{-\Delta_{1}^{a,b}}\,,
\end{aligned}
\end{equation}
where $r_{pq}$ is the distance between the points $p$ and $q$. If $\Delta_{1}^{a,b}$ is large enough, this singularity will lead to a singular behavior of the integral~\eqref{eq:3.2.1} at $r\rightarrow 0$. Thus, singular behavior of correlators leads to the fact that good families for the initial theory cease to be such for the deformed theory.

This entire picture is similar to the geometrical problem about tangent spaces of two close points of a manifold. Ideologically, the set of all quantum field theories can be treated as a nonlinear object, a manifold given by the system of quadratic equations. This system is nothing more than the cutting axiom~\eqref{eq:2.1.1}. Then the tangent space to a given point (a quantum field theory) of the manifold of theories can be considered as the space of local observables of this theory. It is clear that generically the tangent spaces to two different points of the surface do not coincide. 

However, even though $v_{b,r}$ is no longer a good family for the theory $\braket{\,}_{X}^{\left(\mathrm{def}\right)}(g,O_{a})$, we can set the problem of finding a new good family in the form of
\begin{equation}
\label{eq:3.2.3}
\begin{aligned}
\tilde{v}_{b,r}&=v_{b,r}+g\cdot \delta v_{b,r}\,.
\end{aligned}
\end{equation}
Here, $\delta v_{b,r}$ is chosen to eliminate the singular terms in $r$ coming from the integral in~\eqref{eq:3.2.1}. Generally speaking, we could add more terms of the form $g\cdot u$ where $u$ is any local observable for the unperturbed theory, and this would also be a good family for the deformed theory. This kind of freedom in the construction of a new good family is related to the choice of connection on the manifold of quantum field theories. However, when we perform a second deformation of the theory using the local observable $\tilde{v}_{b,r}$, we can eliminate the connection in the second order perturbed partition function by redefining the coupling constants.

In the next section, we calculate $\delta v_{b,r}$ for one important example of the deformation of a two-dimensional conformal field theory. In the occurrence of this correction lies the logarithmic dimension of the observable $\tilde{v}_{b}$ and hence the conformal anomaly.

The fact that there is a non-trivial transformation of the space of local observables as we move along the space of quantum field theories is fundamental. This is where the reasons for the UV divergence in the standard approach to quantum field theory are hidden. In the standard approach, we try to use the local observable of the old theory as a perturbed observable and run into UV divergences during the calculation of loop integrals. Then, we invoke the renormalization procedure. In the functorial formulation, renormalization corresponds to the transport of the tangent vector along the manifold of theories with connection. Exactly this connection is responsible for the conformal anomaly appearance and contains information about the renormalization group.
\section{Beta Function in Two-Dimensional Theories}
\label{sec: Beta Function in Two-Dimensional Theories}
\subsection{Marginal Observables in the Deformed Theory}
\label{subsec: Marginal Observables of Deformed Theory}
Let there be a conformal field theory on surface $X$ with partition function $\braket{\,}_{X}$. We will assume that this theory has a zero central charge (to avoid a non- perturbative contribution to the conformal anomaly), but we expect that our approach can be generalized to the case of a non-zero central charge. We will study its deformations by primary observables. In physics, the deformation of a conformal theory by marginal observable is also known as the conformal perturbation theory \cite{ConfPert1,ConfPert2,ConfPert3,ConfPert4,ConfPert5,ConfPert6}.

Consider the family of all primary marginal observables $\mathcal{M}=\left\{v_{\alpha,r}\right\}_{\alpha=1}^{n}$ of this theory. We will number such observables with Greek indices. In this section, we demonstrate the calculation of the correction $\delta v_{\alpha\beta,r}$ for the observable $v_{\beta,r}$ under the action of deformation by a linear combination of local observables $g^{\alpha}\cdot v_{\alpha,r}$. We will assume that there are no primary observables with negative conformal dimensions in the theory. This requirement is due to the fact that the products of the primary marginal observables would not contain marginal descendants. The details are discussed in appendix~\ref{A}. As noted in~\ref{subsec: Local Observables in Functorial QFT}, when studying local observables, it is sufficient to study them on a flat disk. 

Consider the corresponding deformation on $X=D_{R}$:
\begin{equation}
\label{eq:4.1.1}
\begin{aligned}
\braket{\,}_{D_{R}}^{\left(\mathrm{def}\right)}\left(g,\mathcal{M}\right)&=\braket{\,}_{D_{R}}+g^{\alpha}\cdot \int_{D_{R}}\frac{\mathrm{d}\bar{z}\wedge\mathrm{d}z}{4\pi\mathrm{i}}\,\braket{O_{\alpha}\left(z,\bar{z}\right)}_{D_{R}}\,.
\end{aligned}
\end{equation}
This partition function does not depend on the radius of the disk (it does not depend on the choice of metric on the disk), so in the first order of perturbation theory conformal symmetry is not broken. Now, according to our definition, we need to cut a disk at the point where we want to insert the local observable. Cut out the disk from the center:
\begin{equation}
\label{eq:4.1.2}
\begin{aligned}
\braket{\,}_{D_{R}\backslash D_{r}}^{\left(\mathrm{def}\right)}\left(g,\mathcal{M}\right)&=\braket{\,}_{D_{R}\backslash D_{r}}+g^{\alpha}\cdot \int_{D_{R}\backslash D_{r}}\frac{\mathrm{d}\bar{z}\wedge\mathrm{d}z}{4\pi\mathrm{i}}\,\braket{O_{\alpha}\left(z,\bar{z}\right)}_{D_{R}\backslash D_{r}}\,.
\end{aligned}
\end{equation}
So, we must search for good families for the deformed partition function. Let's look for this family in the form
\begin{equation}
\label{eq:4.1.3}
\begin{aligned}
\tilde{v}_{\beta,r}&=v_{\beta,r}+g^{\alpha}\cdot \delta v_{\alpha\beta,r}\,,
\end{aligned}
\end{equation}
where the family $v_{\beta,r}$ defines the primary marginal observable $v_{\beta}\in\mathcal{M}$. Convolving this family with a partition function $\braket{\,}_{D_{R}\backslash D_{r}}^{\left(\mathrm{def}\right)}\left(g,\mathcal{M}\right)$ we obtain
\begin{equation}
\label{eq:4.1.4}
\begin{aligned}
\left(\braket{\,}_{D_{R}\backslash D_{r}}^{\left(\mathrm{def}\right)}\left(g,\mathcal{M}\right)\otimes \tilde{v}_{\beta,r}\right)_{\partial D_{r}}=\braket{O_{\beta}\left(0\right)}_{D_{R}}+g^{\alpha}\left(\braket{\,}_{D_{R}\backslash D_{r}}\otimes \delta v_{\alpha\beta,r}\right)_{\partial D_{r}}+\\+\,g^{\alpha}\cdot \int_{D_{R}\backslash D_{r}}\frac{\mathrm{d}\bar{z}\wedge\mathrm{d}z}{4\pi\mathrm{i}}\,\braket{O_{\alpha}\left(z,\bar{z}\right)O_{\beta}\left(0\right)}_{D_{R}}\,.
\end{aligned}
\end{equation}
Here we have omitted terms of order $g^2$. For simplicity, we will assume that the OPE of two primary local observables has a special form:\footnote{The calculation for an arbitrary OPE is done in appendix~\ref{A.2}.}
\begin{equation}
\label{eq:4.1.5}
\begin{aligned}
\braket{O_{\alpha}\left(z,\bar{z}\right)O_{\beta}\left(0\right)}_{D_{R}}&=|z|^{-4}\cdot K_{\alpha\beta}^{a}\cdot\braket{O_{a}\left(0\right)}_{D_{R}} +|z|^{-2}\cdot C_{\alpha\beta}^{\,\gamma}\cdot\braket{O_{\gamma}\left(0\right)}_{D_{R}} +\,...
\end{aligned}
\end{equation}
Here, we have written out only the spinless singular part of the OPE since it is the only one that survives after the integration. Local observables numbered with Latin indices are observables with a scaling dimension 0. Then we obtain
\begin{equation}
\label{eq:4.1.6}
\begin{aligned}
 \int_{D_{R}\backslash D_{r}}\frac{\mathrm{d}\bar{z}\wedge\mathrm{d}z}{4\pi\mathrm{i}}\,\braket{O_{\alpha}\left(z,\bar{z}\right)O_{\beta}\left(0\right)}_{D_{R}}=\int_{r}^{R}\rho\,\mathrm{d\rho}\,\rho^{-4}\cdot K_{\alpha\beta}^{a}\cdot\braket{O_{a}\left(0\right)}_{D_{R}}+\\+\int_{r}^{R}\rho\,\mathrm{d\rho}\,\rho^{-2}\cdot C_{\alpha\beta}^{\,\gamma}\cdot\braket{O_{\gamma}\left(0\right)}_{D_{R}}+\mathrm{regular\,part}\,.
 \end{aligned}
\end{equation}
By the regular part here, we mean the result obtained by integrating the regular part of the OPE, which we do not follow. Integration of the first term gives
\begin{equation}
\label{eq:4.1.7}
\begin{aligned}
\int_{r}^{R}\rho\,\mathrm{d}\rho\,\rho^{-4}\cdot K_{\alpha\beta}^{a}\cdot\braket{O_{a}(0)}_{D_{R}}&=\frac{1}{2r^{2}} \cdot K_{\alpha\beta}^{a}\cdot\braket{O_{a}(0)}_{D_{R}} - \frac{1}{2R^{2}}\cdot K_{\alpha\beta}^{a}\cdot\braket{O_{a}(0)}_{D_{R}}\,.
\end{aligned}
\end{equation}
Integration of the second term gives
\begin{equation}
\label{eq:4.1.8}
\begin{aligned}
\int_{r}^{R}\rho\,\mathrm{d}\rho\,\rho^{-2}\cdot C_{\alpha\beta}^{\,\gamma}\cdot\braket{O_{\gamma}(0)}_{D_{R}}&=\log\left(\frac{R}{r}\right)\cdot C_{\alpha\beta}^{\,\gamma}\cdot\braket{O_{\gamma}(0)}_{D_{R}}\,.
\end{aligned}
\end{equation}
We choose the correction $\delta v_{\alpha\beta,r}$ in~\eqref{eq:4.1.3} to eliminate the singular terms in $r$ coming from~\eqref{eq:4.1.7} and~\eqref{eq:4.1.8}. Thus, we obtain the following expression for the correction $\delta v_{\alpha\beta,r}$:
\begin{equation}
\label{eq:4.1.9}
\begin{aligned}
\delta v_{\alpha\beta,r}&=\log(r)\cdot C_{\alpha\beta}^{\,\gamma}\cdot\braket{O_{\gamma}(0)}_{D_{r}} - \frac{1}{2r^{2}}\cdot K_{\alpha\beta}^{a}\cdot\braket{O_{a}(0)}_{D_{r}}\,.
\end{aligned}
\end{equation}

Substituting the calculated family~\eqref{eq:4.1.3} into~\eqref{eq:4.1.2}, we arrive at the correlator of the new local observable $\tilde{O}_{\beta}$ in the deformed theory, evolved from the family $\tilde{v}_{\beta,r}$:
\begin{equation}
\label{eq:4.1.10}
\begin{aligned}
\braket{\tilde{O}_{\beta}(0)}_{D_{R}}^{(\mathrm{def})}\left(g,\mathcal{M}\right) = \braket{O_{\beta}(0)}_{D_{R}} + \log(R)\cdot g^{\alpha}\cdot C_{\alpha\beta}^{\,\gamma}\cdot\braket{O_{\gamma}(0)}_{D_{R}}-\\-\frac{1}{2R^{2}}\cdot g^{\alpha}\cdot K_{\alpha\beta}^{a}\cdot\braket{O_{a}(0)}_{D_{R}}+\mathrm{regular\,part}\,.
\end{aligned}
\end{equation}
Thus, we have calculated one family of local observables for the deformed theory. Now we will be interested in its behavior under scale transformations.
\subsection{Second-Order Beta Function}
\label{subsec: Second-Order Perturbation Theory Beta Function}
We now have a new family of local observables $\tilde{\mathcal{M}}=\left\{\tilde{v}_{\beta,r}\right\}_{\beta=1}^{n}$ in our hands. Let us apply the dilatation operator to the correlator of the new observable $\tilde{O}_{\beta}$ (see appendix~\ref{A.2}):
\begin{equation}
\label{eq:4.2.1}
\begin{aligned}
\lambda^{2}\cdot\mathrm{Dil}_{\lambda} \braket{\tilde{O}_{\beta}(0)}_{D_{R}}^{(\mathrm{def})}(g,\mathcal{M})=\lambda^{2}\cdot\braket{\tilde{O}_{\beta}(0)}_{D_{\lambda\cdot R}}^{(\mathrm{def})}(g,\mathcal{M})=\\=\braket{\tilde{O}_{\beta}(0)}_{D_{R}}^{(\mathrm{def})}(g,\mathcal{M})+\log(\lambda)\cdot g^{\alpha}\cdot C_{\alpha\beta}^{\,\gamma}\cdot\braket{O_{\gamma}(0)}_{D_{R}}\,.
\end{aligned}
\end{equation}
Thus, we see that the local observable has acquired a logarithmic (anomalous) dimension.\footnote{Note that the formula~\eqref{eq:4.2.1} can be rewritten in terms of good families as $\lambda^{2}\cdot\mathrm{Dil}_{\lambda} \tilde{v}_{\beta,r}=\tilde{v}_{\beta,r}+\log(\lambda)\cdot g^{\alpha}\cdot C_{\alpha\beta}^{\,\gamma}\cdot\tilde{v}_{\gamma,r}$. Thus, the dilatation operator still acts inside the space of local observables of the deformed theory.} The reason for the non-trivial transformation is the presence of a logarithm in the first summand of~\eqref{eq:4.1.9}. This term contains the structure constants responsible for the marginal sector of the operator algebra.

Using the correlator of the local observable $\tilde{v}_{\beta,r}$ we can construct the partition function in the second order of perturbation theory. The deformed correlation function has the form:
\begin{equation}
\label{eq:4.2.1.0.0}
\begin{aligned}
\braket{\tilde{O}_{\alpha}(z,\bar{z})}_{D_{R}}^{(\mathrm{def})}\left(g,\mathcal{M}\right)=\braket{O_{\alpha}\left(z,\bar{z}\right)}_{D_{R}}\,+\\+\,g^{\alpha}\cdot \mathcal{C}^{\,\gamma}_{\alpha\beta}(z,\bar{z};R)\cdot \braket{O_{\gamma}\left(z,\bar{z}\right)}_{D_{R}}+g^{\alpha}\cdot \mathcal{K}_{\alpha\beta}^{\,c}\left(z,\bar{z};R\right)\cdot\braket{O_{c}\left(z,\bar{z}\right)}_{D_{R}}+...\,
\end{aligned}
\end{equation}
This formula is discussed in appendix~\ref{A.3}. Here, $\mathcal{C}^{\,\gamma}_{\alpha\beta}(z,\bar{z};R)$ and $\mathcal{K}_{\alpha\beta}^{\,c}\left(z,\bar{z};R\right)$ are functions on the disk $D_{R}$. To demonstrate the emergence of a conformal anomaly, it is sufficient to consider only the first coefficient in the series for $\mathcal{C}^{\,\gamma}_{\alpha\beta}(z,\bar{z};R)$:
\begin{equation}
\label{eq:4.2.1.0.1}
\begin{aligned}
\mathcal{C}^{\,\gamma}_{\alpha\beta}(z,\bar{z};R)=\mathcal{C}^{\,\gamma}_{\alpha\beta}(0,0;R)+...=C_{\alpha\beta}^{\,\gamma}\cdot \mathrm{log}(R)+...\,
\end{aligned}
\end{equation}

We construct the second-order perturbative partition function via double deformation. That is, according to the definition of a deformation given in section~\ref{subsec: Deformation of the Partition Function}, we construct the deformation of the theory $\braket{\,}_{D_{R}}^{(\mathrm{def})}$ by the family of local observables $\tilde{\mathcal{M}}$:
\begin{equation}
\label{eq:4.2.1.0}
\begin{aligned}
\braket{\,}_{D_{R}}^{\left(2-\mathrm{def}\right)}(g, \mathcal{M};\tilde{g},\tilde{\mathcal{M}})&=\braket{\,}_{D_{R}}^{\left(\mathrm{def}\right)}(g, \mathcal{M})+\tilde{g}^{\alpha}\cdot \int_{D_{R}}\frac{\mathrm{d}\bar{z}\wedge\mathrm{d}z}{4\pi\mathrm{i}}\,\braket{\tilde{O}_{\alpha}(z,\bar{z})}_{D_{R}}^{(\mathrm{def})}\left(g,\mathcal{M}\right)\,.
\end{aligned}
\end{equation}
This partition function satisfies the cutting axiom by construction. When constructing the deformation, we assume that $g^{\alpha}\cdot g^{\beta}=0$ and $\tilde{g}^{\alpha}\cdot \tilde{g}^{\beta}=0$. Let's rewrite it in more detail:
\begin{equation}
\label{eq:details}
\begin{aligned}
\braket{\,}_{D_{R}}^{\left(2-\mathrm{def}\right)}=\braket{\,}_{D_{R}}+\left(g^{\alpha}+\tilde{g}^{\alpha}\right)\cdot \int_{D_{R}}\frac{\mathrm{d}\bar{z}\wedge\mathrm{d}z}{4\pi\mathrm{i}}\,\braket{O_{\alpha}\left(z,\bar{z}\right)}_{D_{R}}+\\+\,g^{\alpha}\tilde{g}^{\beta}\cdot \int_{D_{R}}\frac{\mathrm{d}\bar{z}\wedge\mathrm{d}z}{4\pi\mathrm{i}}\,\left\{\log(R)\cdot C_{\alpha\beta}^{\,\gamma}\cdot\braket{O_{\gamma}(z,\bar{z})}_{D_{R}}+\,...\right\}\,.
\end{aligned}
\end{equation}
Here we have omitted the part of the correlator~\eqref{eq:4.2.1.0.0} that does not contribute to the conformal anomaly. Now, note that for a symmetric symbol $S_{\alpha\beta}$ and first order nilpotents $g^{\alpha},\tilde{g}^{\beta}$, it is true that
\begin{equation}
\label{eq:coupling.2}
\begin{aligned}
\tilde{g}^{\alpha}g^{\beta}\cdot S_{\alpha\beta}&=\frac{(\tilde{g}^{\alpha}+g^{\alpha})(\tilde{g}^{\beta}+g^{\beta})}{2}\cdot S_{\alpha\beta}\,.
\end{aligned}
\end{equation} 
Then the partition function~\eqref{eq:details} actually depends only on the sum
\begin{equation}
\label{eq:coupling.1}
\begin{aligned}
g_{c}^{\alpha}&=g^{\alpha}+\tilde{g}^{\alpha}\,.
\end{aligned}
\end{equation}
We will denote it by $g_{c}^{\alpha}$, the index \(c\) meaning ‘‘coupling’’. This sum is the coupling constant of the perturbative theory in the second order. Such a sum of first order nilpotents is a second-order nilpotent.  It turns out that perturbation theory of any order can be constructed similarly via multiple deformations. For a detailed explanation of how multiple deformation is related to perturbation theory, we refer to \cite{LosevGritskov}. In addition, in appendix~\ref{C.2} we discuss a simple example of constructing a perturbative partition function in second order via double deformation. Thus, we obtain
\begin{equation}
\label{eq:4.2.2.1}
\begin{aligned}
\braket{\,}_{D_{R}}^{\left(2-\mathrm{def}\right)}=\braket{\,}_{D_{R}}+g_{c}^{\alpha}\cdot \int_{D_{R}}\frac{\mathrm{d}\bar{z}\wedge\mathrm{d}z}{4\pi\mathrm{i}}\,\braket{O_{\alpha}\left(z,\bar{z}\right)}_{D_{R}}+\\+\,\frac{g_{c}^{\alpha}g_{c}^{\beta}}{2}\cdot \int_{D_{R}}\frac{\mathrm{d}\bar{z}\wedge\mathrm{d}z}{4\pi\mathrm{i}}\,\left\{\log(R)\cdot C_{\alpha\beta}^{\,\gamma}\cdot\braket{O_{\gamma}(z,\bar{z})}_{D_{R}}+\,...\right\}\,.
\end{aligned}
\end{equation}
Then such a partition function starts to depend on the radius of the disk:
\begin{equation}
\label{eq:4.2.3}
\begin{aligned}
\braket{\,}_{D_{\lambda\cdot R}}^{\left(2-\mathrm{def}\right)}&=\braket{\,}_{D_{R}}^{\left(2-\mathrm{def}\right)}+\log(\lambda)\cdot\frac{g_{c}^{\alpha}g_{c}^{\beta}}{2}\cdot C_{\alpha\beta}^{\,\gamma}\cdot\int_{D_{R}}\frac{\mathrm{d}\bar{z}\wedge\mathrm{d}z}{4\pi\mathrm{i}}\,\left\{\braket{O_{\gamma}(z,\bar{z})}_{D_{R}}\right\}\,.
\end{aligned}
\end{equation}
This is the well-known equation that defines the perturbative conformal anomaly \cite{ZZ}.

The radius scaling is a Weyl transformation of the metric, which can be carried to the perturbative coordinate chart of coupling constants. In other words, this transformation can be simulated by the coupling constants transformation:
\begin{equation}
\label{eq:4.2.4}
\begin{aligned}
g^{\gamma}_{c}\rightarrow g_{c}^{\gamma}(\lambda)&=g_{c}^{\gamma}+\log(\lambda)\cdot \frac{g_{c}^{\alpha}g_{c}^{\beta}}{2}\cdot C_{\alpha\beta}^{\,\gamma}\,.
\end{aligned}
\end{equation}
This is the defining relation for running coupling constants. Thus the beta function
\begin{equation}
\label{eq:4.2.5}
\begin{aligned}
\beta^{\gamma}(g_{c})=\frac{\mathrm{d}g_{c}^{\gamma}(\lambda)}{\mathrm{d}\log(\lambda)}&=\frac{g_{c}^{\alpha}g_{c}^{\beta}}{2}\cdot C_{\alpha\beta}^{\,\gamma}\,,
\end{aligned}
\end{equation}
which agrees with the known result for the one-loop beta function in two-dimensional theories \cite{Zamolodchikov:1987ti}.

We have demonstrated the dependence of the partition function of the perturbed theory on the conformal factor of the metric. The reason for the anomaly lies in the logarithmic dimension, which the marginal observables acquire after the first deformation. In turn, the anomalous dimension reflects the presence of a non-zero connection on the space of theories, since it arises as a result of deformation of the space of local observables.
\section{Conclusions}
\label{sec: Conclusion and Discussion}
In this paper, we constructed the second order beta function in FQFT. The described
deformation procedure directly generalizes to higher orders of perturbation theory. In this approach we are not burdened with the need to introduce a single regularization prescription for all orders of perturbation theory, everything turns out to be natural. 

From different points of view, it would be interesting to study the functorial beta function and its properties in higher orders of perturbation theory. Based on the results obtained in \cite{LosevShifmanGamayun, LosevShifmanGamayun2}, it has been conjectured that the first coefficients of the beta function can be interpreted as algebraic operations in some $A_{\infty}$-algebra. Therefore, the authors expect that the approach described here will help develop a conceptual view of the algebraic structure of the beta function. 

Another interesting question is the consideration of contact terms in correlation functions. At first sight, the definition of local observables and their OPE in FQFT is unable to account for singularities with support at a one point. In other approaches, the consideration of contact terms leads to important phenomena, such as the renormalization of the stress energy tensor \cite{Cardy:1994pb}. Strictly speaking, this question remains open, but there is hope that the appreciable effects related to contact terms in the Feynman approach have analogues in the functorial approach.

\section{Acknowledgements}
We are grateful to Alexey Litvinov, Nicolai Reshetikhin, Oleksander Gamayun, Tim Sulimov and Vyacheslav Rychkov for helpful discussions. 

We are grateful to the Mittag-Leffler Institute in Djursholm for providing a workspace for the duration of the research program Cohomological Aspects of QFT 2025. 

The first author is supported by the Ministry of Science and Higher Education of the Russian Federation (Agreement
No. 075-15-2025-013).

The first author is a winner of the “Leader” contest conducted by the Foundation
for the Advancement of Theoretical Physics and Mathematics “BASIS” and would
like to thank its sponsors and jury.

\appendix
\section{Calculation for an Arbitrary OPE}
\label{A}
\subsection{General OPE Formula for Marginal Primaries}
\label{A.1}
The purpose of this appendix is to generalize formula~\eqref{eq:4.1.9} to the case of arbitrary conformal operator algebras. We will consider CFT on a flat disk $D_{R}$. The space of boundary conditions in functorial CFT is a representation of the Virasoro algebra. We work under the assumption that there are no primary fields of negative integer dimensions whose descendants could behave as quasi-primary marginal observables.

Consider the set of all primary observables $\mathcal{P}$ of the theory given by the partition function $\braket{\,}_{D_{R}}$. The set of marginal primaries will be denoted as $\mathcal{M}$. There is a formula for the OPE of two primary observables \cite{BPZ, MnevLect}:
\begin{equation}
\begin{aligned}
\braket{O_{a}\left(z,\bar{z}\right)O_{b}\left(0\right)}_{D_{R}}&=\sum_{c,\left\{\mu\right\},\left\{\bar{\mu}\right\}} C_{ab}^{\,c}\left(\left\{\mu\right\},\left\{\bar{\mu}\right\}\right)\cdot z^{\Delta_{abc}(\left\{\mu\right\})}\cdot \bar{z}^{\bar{\Delta}_{abc}(\left\{\bar{\mu}\right\})}\cdot\braket{O_{c}^{\left\{\mu\right\},\left\{\bar{\mu}\right\}}\left(0\right)}_{D_{R}}\,,
\\
\Delta_{abc}(\left\{\mu\right\})&=h_{c}+|\left\{\mu\right\}|-h_{a}-h_{b}\,,
\qquad
\bar{\Delta}_{abc}(\left\{\bar{\mu}\right\})=\bar{h}_{c}+|\left\{\bar{\mu}\right\}|-\bar{h}_{a}-\bar{h}_{b}\,.
\end{aligned}
\end{equation}
Here, we have introduced the holomorphic $h\geq 0$ and antiholomorphic $\bar{h}\geq 0$ dimensions, so that the conformal dimension $\Delta=h+\bar{h}$. The coefficients $C_{ab}^{\,c}\left(\left\{\mu\right\}, \left\{\bar{\mu}\right\}\right)$ are conformal theory data. The information about them is contained in the spectrum of the theory. Clarify the notation: $|\left\{\mu\right\}|$ is the sum $\mu_{1}+...+\mu_{k}$ of partition elements. In our convention, the elements of $\left\{\mu\right\}$ are ordered as $\mu_{1}\geq \mu_{2}\geq ...\geq\mu_{k}>0$. As usual, the boundary state spaces are products of the holomorphic and antiholomorphic representations of the Virasoro algebra. Therefore, there are two sets of generators $L_{n}$ and $\bar{L}_{n}$ acting in the space of boundary conditions. Let's consider a good family
\begin{equation}
\label{eq:A.1.1}
\begin{aligned}
v_{a,r}&=\braket{O_{a}\left(0\right)}_{D_{r}}
\end{aligned}
\end{equation}
defining some primary local observable with dimension $\Delta_{a}=h_{a}+\bar{h}_{a}$. Then good families corresponding to conformal descendants are given by the formula:
\begin{equation}
\label{eq:A.1.2}
\begin{aligned}
\braket{O_{a}^{\left\{\mu\right\},\left\{\bar{\mu}\right\}}\left(0\right)}_{D_{r}}&=r^{-L_{0}-\bar{L}_{0}}\,\left(L_{-\mu_{1}}\,L_{-\mu_{2}}...\right)\left(\bar{L}_{-\bar{\mu}_{1}}\,\bar{L}_{-\bar{\mu}_{2}}...\right)\,r^{L_{0}+\bar{L}_{0}}\braket{O_{a}\left(0\right)}_{D_{r}}\,.
\end{aligned}
\end{equation}
It is clear how the dilatation operator acts on the descendant:
\begin{equation}
\label{eq:A.1.3}
\begin{aligned}
\mathrm{Dil}_{\lambda} \braket{O_{a}^{\left\{\mu\right\},\left\{\bar{\mu}\right\}}\left(0\right)}_{D_{r}}&=\lambda^{-h_{a}-\bar{h}_{a}-|\left\{\mu\right\}|-|\left\{\bar{\mu}\right\}|}\cdot\braket{O_{a}^{\left\{\mu\right\},\left\{\bar{\mu}\right\}}\left(0\right)}_{D_{r}}\,.
\end{aligned}
\end{equation}

It is necessary to study the general OPE formula for the case where $O_{a},O_{b}\in\mathcal{M}$:
\begin{equation}
\begin{aligned}
\braket{O_{a}\left(z,\bar{z}\right)O_{b}\left(0\right)}_{D_{R}}&=\sum_{c,\left\{\mu\right\},\left\{\bar{\mu}\right\}} C_{ab}^{\,c}\left(\left\{\mu\right\},\left\{\bar{\mu}\right\}\right)\cdot z^{\delta_{c}^{\left\{\mu\right\}}}\cdot \bar{z}^{\bar{\delta}_{c}^{\left\{\bar{\mu}\right\}}}\cdot\braket{O_{c}^{\left\{\mu\right\},\left\{\bar{\mu}\right\}}\left(0\right)}_{D_{R}}\,,
\\
\delta_{c}^{\left\{\mu\right\}}&=h_{c}+|\left\{\mu\right\}|-2\,,\qquad \bar{\delta}_{c}^{\left\{\bar{\mu}\right\}}=\bar{h}_{c}+|\left\{\bar{\mu}\right\}|-2\,.
\end{aligned}
\end{equation}
Let us separate from this sum the terms where the correlation functions of the marginal observables stand. There are such summands when $\left\{\mu\right\}=\left\{\bar{\mu}\right\}=\left\{1\right\}$ and $h_{c}=\bar{h}_{c}=0$ or when $\left\{\mu\right\}=\left\{\bar{\mu}\right\}=\left\{\emptyset\right\}$ and $h_{c}=\bar{h}_{c}=1$.\footnote{Generally speaking, there may be situations when in the theory there are spin primary observables of dimensions \( (0,1) \) and  \( (1,0) \) and then the summands with \( \left\{\mu\right\} = \left\{1\right\}, \left\{\bar{\mu}\right\} = \left\{0\right\}\) or \( \left\{\mu\right\} = \left\{0\right\}, \left\{\bar{\mu}\right\} = \left\{1\right\}\) will also contribute. However, it is easy to see that the observables \( \braket{O_{c}^{\left\{1\right\},\left\{0\right\}}\left(0\right)}_{D_{r}} \) and \( \braket{O_{c}^{\left\{0\right\},\left\{1\right\}}\left(0\right)}_{D_{r}}\) are also primary marignals, so in the presence of such observables with spin the calculation is similar.} It can be proven that the descendant family $\braket{O_{a}^{\left\{1\right\},\left\{1\right\}}\left(0\right)}_{D_{r}}$ is actually primary if the observable $\braket{O_{a}\left(0\right)}_{D_{r}}$ had conformal dimensions $h_{a}=\bar{h}_{a}=0$. The easiest way to be sure of this is to check that the boundary state
\begin{equation}
\label{eq:A.1.4}
\begin{aligned}
\braket{O_{a}^{\left\{1\right\},\left\{1\right\}}\left(0\right)}_{D_{r}}&=r^{-L_{0}-\bar{L}_{0}}\,L_{-1}\bar{L}_{-1}\,r^{L_{0}+\bar{L}_{0}}\braket{O_{a}\left(0\right)}_{D_{r}}
\end{aligned}
\end{equation}
is a singular vector. But then it is a combination of primary marginal:
\begin{equation}
\label{eq:A.1.5}
\begin{aligned}
\braket{O_{a}^{\left\{1\right\},\left\{1\right\}}\left(0\right)}_{D_{R}}&=M^{b}_{a}\cdot \braket{O_{b}^{\left\{0\right\},\left\{0\right\}}\left(0\right)}_{D_{R}}=M^{b}_{a}\cdot \braket{O_{b}\left(0\right)}_{D_{R}}\,.
\end{aligned}
\end{equation}
We will now number primary marginals with Greek symbols:
\begin{equation}
\begin{aligned}
\braket{O_{\alpha}\left(z,\bar{z}\right)O_{\beta}\left(0\right)}_{D_{R}}=\left(C_{\alpha\beta}^{\,\gamma}\left(\left\{\emptyset\right\},\left\{\emptyset\right\}\right)+C_{\alpha\beta}^{\,a}\left(\left\{1\right\},\left\{1\right\}\right)\cdot M_{a}^{\gamma}\right)\cdot|z|^{-2}\cdot\braket{O_{\gamma}\left(0\right)}_{D_{R}}+\\+\sum^{\sim}_{c,\left\{\mu\right\},\left\{\bar{\mu}\right\}} C_{\alpha\beta}^{\,c}\left(\left\{\mu\right\},\left\{\bar{\mu}\right\}\right)\cdot z^{\left(h_{c}+|\left\{\mu\right\}|-2\right)}\cdot \bar{z}^{\left(\bar{h}_{c}+|\left\{\bar{\mu}\right\}|-2\right)}\cdot\braket{O_{c}^{\left\{\mu\right\},\left\{\bar{\mu}\right\}}\left(0\right)}_{D_{R}}\,.
\end{aligned}
\end{equation}
In this sum, the index \(c\) still runs throughout the set, numbering all primary observables $\mathcal{P}$, while $\gamma$ runs from $1$ to $n$, enumerating all marginal primaries from $\mathcal{M}=\left\{O_{\alpha}\right\}_{\alpha=1}^{n}$. The sum with a tilde means the sum over all primaries with the corresponding marginal terms removed. To harmonize the notation, we will introduce a new one:
\begin{equation}
\label{eq:A.1.6}
\begin{aligned}
C_{\alpha\beta}^{\,\gamma}&=C_{\alpha\beta}^{\,\gamma}\left(\left\{\emptyset\right\},\left\{\emptyset\right\}\right)+C_{\alpha\beta}^{\,a}\left(\left\{1\right\},\left\{1\right\}\right)\cdot M_{a}^{\gamma}\,.
\end{aligned}
\end{equation}
We are now ready to generalize the result of section~\ref{subsec: Marginal Observables of Deformed Theory} on the deformation of marginal families arising by perturbation of the initial conformal theory.
\subsection{Deformation of Marginal Primaries in the General Case}
\label{A.2}
First, briefly recall what we computed in section~\ref{subsec: Marginal Observables of Deformed Theory}. Deforming the conformal field theory by the marginal observable $g^{\alpha}\cdot v_{\alpha,r}$, we observe that the old good families are no longer such. We search for a correction of the family that generates the marginal observable $v_{\beta,r}$. 

Consider the corresponding deformed theory on a disk $D_{R}$:
\begin{equation}
\label{eq:A.2.1}
\begin{aligned}
\braket{\,}_{D_{R}}^{\left(\mathrm{def}\right)}\left(g,\mathcal{M}\right)&=\braket{\,}_{D_{R}}+g^{\alpha}\cdot \int_{D_{R}}\frac{\mathrm{d}\bar{z}\wedge\mathrm{d}z}{4\pi\mathrm{i}}\,\braket{O_{\alpha}\left(z,\bar{z}\right)}_{D_{R}}\,.
\end{aligned}
\end{equation}
By cutting a small disk from the center we obtain
\begin{equation}
\label{eq:A.2.2}
\begin{aligned}
\braket{\,}_{D_{R}\backslash D_{r}}^{\left(\mathrm{def}\right)}\left(g,\mathcal{M}\right)&=\braket{\,}_{D_{R}\backslash D_{r}}+g^{\alpha}\cdot \int_{D_{R}\backslash D_{r}}\frac{\mathrm{d}\bar{z}\wedge\mathrm{d}z}{4\pi\mathrm{i}}\,\braket{O_{\alpha}\left(z,\bar{z}\right)}_{D_{R}\backslash D_{r}}\,.
\end{aligned}
\end{equation}
Now let's try to insert there a family of the form
\begin{equation}
\label{eq:A.2.3}
\begin{aligned}
\tilde{v}_{\beta,r}&=v_{\beta,r}+g^{\alpha}\cdot C_{\alpha}^{b}\cdot w_{b,r}+g^{\alpha}\cdot\delta v_{\alpha\beta,r}\,,
\end{aligned}
\end{equation}
where family $v_{\beta,r}\in\mathcal{M}$ define primary marginals and $C^{b}_{\alpha}\cdot w_{b,r}$ define any finite linear combination of local observables in non-deformed theory. Now we convolute this family with a partition function $\braket{\,}_{D_{R}\backslash D_{r}}^{\left(\mathrm{def}\right)}\left(g,\mathcal{M}\right)$ and obtain
\begin{equation}
\label{eq:A.2.4}
\begin{aligned}
\left(\braket{\,}_{D_{R}\backslash D_{r}}^{\left(\mathrm{def}\right)}\left(g\right)\otimes \tilde{v}_{\beta,r}\right)_{\partial D_{r}}=\braket{O_{\beta}\left(0\right)}_{D_{R}}+g^{\alpha}\cdot C_{\alpha}^{b}\cdot\braket{O_{b}\left(0\right)}_{D_{R}} +\\+\,g^{\alpha}\cdot\left(\braket{\,}_{D_{R}\backslash D_{r}}\otimes \delta v_{\alpha\beta,r}\right)_{\partial D_{r}}+g^{\alpha}\cdot \int_{D_{R}\backslash D_{r}}\frac{\mathrm{d}\bar{z}\wedge\mathrm{d}z}{4\pi\mathrm{i}}\,\braket{O_{\alpha}\left(z,\bar{z}\right)O_{\beta}\left(0\right)}_{D_{R}}\,.
\end{aligned}
\end{equation}

Now we need to compute the integral:
\begin{equation}
\label{eq:A.2.5}
\begin{aligned}
\int_{D_{R}\backslash D_{r}}\frac{\mathrm{d}z\wedge\mathrm{d}\bar{z}}{4\pi\mathrm{i}}\,\braket{O_{\alpha}\left(z,\bar{z}\right)O_{\beta}\left(0\right)}_{D_{R}}=\int_{D_{R}\backslash D_{r}}\frac{\mathrm{d}\bar{z}\wedge\mathrm{d}z}{4\pi\mathrm{i}}\,\left\{C_{\alpha\beta}^{\,\gamma}\cdot|z|^{-2}\cdot\braket{O_{\gamma}\left(0\right)}_{D_{R}}\right\}+\\+\sum^{\sim}_{c,\left\{\mu\right\},\left\{\bar{\mu}\right\}} C_{\alpha\beta}^{\,c}\left(\left\{\mu\right\},\left\{\bar{\mu}\right\}\right)\cdot\braket{O_{c}^{\left\{\mu\right\},\left\{\bar{\mu}\right\}}\left(0\right)}_{D_{R}}\cdot \int_{D_{R}\backslash D_{r}}\frac{\mathrm{d}\bar{z}\wedge\mathrm{d}z}{4\pi\mathrm{i}}\,z^{\left(h_{c}+|\left\{\mu\right\}|-2\right)}\cdot \bar{z}^{\left(\bar{h}_{c}+|\left\{\bar{\mu}\right\}|-2\right)}\,.
\end{aligned}
\end{equation}
The first integral will give the logarithmic contribution. It is clear that in the second integral only those summands in which $h_{c}+|\left\{\mu\right\}|=\bar{h}_{c}+|\left\{\bar{\mu}\right\}|$. Then we get
\begin{equation}
\label{eq:A.2.6}
\begin{aligned}
\sum^{\sim}_{c,\left\{\mu\right\},\left\{\bar{\mu}\right\}} C_{\alpha\beta}^{\,c}\left(\left\{\mu\right\},\left\{\bar{\mu}\right\}\right)\cdot\braket{O_{c}^{\left\{\mu\right\},\left\{\bar{\mu}\right\}}\left(0\right)}_{D_{R}}\cdot \int_{D_{R}\backslash D_{r}}\frac{\mathrm{d}\bar{z}\wedge\mathrm{d}z}{4\pi\mathrm{i}}\,z^{\left(h_{c}+|\left\{\mu\right\}|-2\right)}\cdot \bar{z}^{\left(\bar{h}_{c}+|\left\{\bar{\mu}\right\}|-2\right)}=\\=\sum_{h_{c}+|\left\{\mu\right\}|=\bar{h}_{c}+|\left\{\bar{\mu}\right\}|\neq 1} C_{\alpha\beta}^{\,c}\left(\left\{\mu\right\},\left\{\bar{\mu}\right\}\right)\cdot\braket{O_{c}^{\left\{\mu\right\},\left\{\bar{\mu}\right\}}\left(0\right)}_{D_{R}}\cdot \frac{R^{2\left(h_{c}+|\left\{\mu\right\}|-1\right)}-r^{2\left(h_{c}+|\left\{\mu\right\}|-1\right)}}{2\left(h_{c}+|\left\{\mu\right\}|-1\right)}\,.
\end{aligned}
\end{equation}
Thus we obtained a formula for a good family $\tilde{v}_{\beta,r}$ defining the local observable of the deformed theory $\braket{\,}_{D_{R}}^{\left(\mathrm{def}\right)}\left(g,\mathcal{M}\right)$:
\begin{equation}
\label{eq:A.2.7}
\begin{aligned}
\tilde{v}_{\beta,r}=v_{\beta,r}+g^{\alpha}\cdot C_{\alpha}^{b}\cdot w_{b,r}+g_{\alpha}\cdot C^{\,\gamma}_{\alpha\beta}\cdot \braket{O_{\gamma}\left(0\right)}_{D_{r}}\cdot \log\left(r\right)\,+\\+\sum_{h_{c}+|\left\{\mu\right\}|=\bar{h}_{c}+|\left\{\bar{\mu}\right\}|\neq 1} g^{\alpha}\cdot C_{\alpha\beta}^{\,c}\left(\left\{\mu\right\},\left\{\bar{\mu}\right\}\right)\cdot\braket{O_{c}^{\left\{\mu\right\},\left\{\bar{\mu}\right\}}\left(0\right)}_{D_{r}}\cdot \frac{r^{2\left(h_{c}+|\left\{\mu\right\}|-1\right)}}{2\left(h_{c}+|\left\{\mu\right\}|-1\right)}\,.
\end{aligned}
\end{equation}

Let us now make a final comment about scaling. In section~\ref{subsec: Second-Order Perturbation Theory Beta Function} We have written a formula~\eqref{eq:4.2.1} for scaling law of the deformed observable $\tilde{v}_{\beta,r}$ when $w_{b,r}=0$. Let us derive it. First, we study how the summands from the last sum in~\eqref{eq:A.2.7} are scaled:
\begin{equation}
\label{eq:A.2.8}
\begin{aligned}
\mathrm{Dil}_{\lambda}\left[r^{2\left(h_{c}+|\left\{\mu\right\}|-1\right)}\cdot\braket{O_{c}^{\left\{\mu\right\},\left\{\bar{\mu}\right\}}\left(0\right)}_{D_{r}}\right]&=\lambda^{d-2}\cdot r^{2\left(h_{c}+|\left\{\mu\right\}|-1\right)}\cdot\braket{O_{c}^{\left\{\mu\right\},\left\{\bar{\mu}\right\}}\left(0\right)}_{D_{r}}\,,
\end{aligned}
\end{equation}
where $d=h_{c}-\bar{h}_{c}+|\left\{\mu\right\}|-|\left\{\bar{\mu}\right\}|$. But the sum is over the set where $h_{c}+|\left\{\mu\right\}|=\bar{h}_{c}+|\left\{\bar{\mu}\right\}|$, then $d=0$ and we obtain that
\begin{equation}
\label{eq:A.2.9}
\begin{aligned}
\mathrm{Dil}_{\lambda}\left[r^{2\left(h_{c}+|\left\{\mu\right\}|-1\right)}\cdot\braket{O_{c}^{\left\{\mu\right\},\left\{\bar{\mu}\right\}}\left(0\right)}_{D_{r}}\right]&=\lambda^{-2}\cdot r^{2\left(h_{c}+|\left\{\mu\right\}|-1\right)}\cdot\braket{O_{c}^{\left\{\mu\right\},\left\{\bar{\mu}\right\}}\left(0\right)}_{D_{r}}\,.
\end{aligned}
\end{equation}
Then only the third summand that contains the structure constants of the marginal OPE sector contributes to the logarithmic dimension of $\tilde{v}_{\beta,r}$. Now it is easy to see that
\begin{equation}
\label{eq:A.2.10}
\begin{aligned}
\lambda^{2}\cdot\mathrm{Dil}_{\lambda} \tilde{v}_{\beta,r}&=\tilde{v}_{\beta,r}+\log(\lambda)\cdot g^{\alpha}\cdot C_{\alpha\beta}^{\,\gamma}\cdot\braket{O_{\gamma}(0)}_{D_{r}}\,.
\end{aligned}
\end{equation}
\subsection{Deformed Correlation Function on a Disk}
\label{A.3}
We now compute the deformed correlation function on the disk $D_{R}$. By definition, to compute the deformed correlator of the local observable at the point $(z,\bar{z})$ we should cut the disk around this point and insert the deformed good family we computed in the previous section. By definition
\begin{equation}
\label{eq:A.3.1}
\begin{aligned}
\braket{\tilde{O}_{\beta}(z,\bar{z})}^{(\mathrm{def})}_{D_{R}}(g,\mathcal{M})=\lim_{r\rightarrow 0}\left\{\left(\braket{\,}^{(\mathrm{def})}_{D_{R}\backslash D_{r}(z)}(g,\mathcal{M})\otimes \braket{\tilde{O}_{\beta}(0)}^{(\mathrm{def})}_{D_{r}}(g,\mathcal{M})\right)_{\partial D_{r}(z)}\right\}\,.
\end{aligned}
\end{equation}
Then we obtain the following result for the deformed correlation function on the disk:
\begin{equation}
\label{eq:A.3.2}
\begin{aligned}
    \braket{\tilde{O}_{\beta}(z,\bar{z})}^{(\mathrm{def})}_{D_{R}}(g,\mathcal{M})=\braket{O_{\beta}(z,\bar{z})}_{D_{R}}+\sum_{c,\{\mu\},\{\bar{\mu}\}}\mathcal{F}_{\alpha\beta}^{\,c}(z,\bar{z};\{\mu\},\{\bar{\mu}\};R)\cdot\braket{O_{c}^{\{\mu\},\{\bar{\mu}\}}(z,\bar{z})}_{D_{R}}\,,
\end{aligned}
\end{equation}
where $\mathcal{F}_{ab}^{c}(z,\bar{z};\{\mu\},\{\bar{\mu}\};R)$ is given by the $r\rightarrow 0$ limit of
\begin{equation}
\label{eq:A.3.3}
\begin{aligned}
    \mathcal{F}_{\alpha\beta}^{\,c}(z,\bar{z};\{\mu\},\{\bar{\mu}\};R,r)=\delta_{h_{c}+|\left\{\mu\right\}|,\bar{h}_{c}+|\left\{\bar{\mu}\right\}|}\cdot C_{\alpha\beta}^{\,c}(\{\mu\},\{\bar{\mu}\})\cdot\frac{r^{2(h_{c}+|\left\{\mu\right\}|-1)}}{2(h_{c}+|\left\{\mu\right\}|-1)}\,+\\+\,C_{\alpha\beta}^{\,c}(\{\mu\},\{\bar{\mu}\})\cdot\int_{D_{R}\backslash D_{r}(z)}\frac{\mathrm{d}\bar{w}\wedge\mathrm{d}w}{4\pi\mathrm{i}}\,\frac{1}{(w-z)^{2-h_{c}-|\{\mu\}|}\cdot (\bar{w}-\bar{z})^{2-\bar{h}_{c}-|\{\bar{\mu}\}|}}\,.
\end{aligned}
\end{equation}
Then $\mathcal{C}_{\alpha\beta}^{\,\gamma}(z,\bar{z},R)$ mentioned in~\eqref{eq:4.2.1.0.0} are expressed via $\mathcal{F}_{\alpha\beta}^{\,\gamma}(z,\bar{z};\{\mu\},\{\bar{\mu}\};R)$ as
\begin{equation}
\label{eq:A.3.4}
\begin{aligned}
    \mathcal{C}_{\alpha\beta}^{\,\gamma}(z,\bar{z},R)=\mathcal{F}_{\alpha\beta}^{\,\gamma}(z,\bar{z};\{\emptyset\},\{\emptyset\};R)\,.
\end{aligned}
\end{equation}

When $z=\bar{z}=0$, only the terms in which $h_{c}+|\left\{\mu\right\}|=\bar{h}_{c}+|\left\{\bar{\mu}\right\}|$ survive in~\eqref{eq:A.3.2}, it will be equal to~\eqref{eq:A.2.7}. And hence
\begin{equation}
\label{eq:A.3.5}
\begin{aligned}
    \mathcal{C}_{\alpha\beta}^{\,\gamma}(z,\bar{z},R)=\mathcal{C}_{\alpha\beta}^{\,\gamma}(0,0,R)+...=C_{\alpha\beta}^{\,\gamma}\cdot\log(R)+...
\end{aligned}
\end{equation}
\section{Example I: Functorial Formulation of Quantum Mechanics}
\label{C}
\subsection{Quantum Mechanics as One-Dimensional FQFT}
\label{C.1}
In this appendix we will formulate quantum mechanics as one-dimensional FQFT. Then we will demonstrate how second-order perturbation theory is obtained from the double deformation using a quantum mechanics example.

As a manifold $X$, we consider the line segment $[\alpha,\beta]$. The boundary of this segment is two points: $\partial X=\partial \,[\alpha,\beta]= \left\{\alpha\right\}\sqcup \left\{\beta\right\}$. Then $\mathcal{H}_{\{\alpha\}}=\mathcal{H}^{\,*}$ and $\mathcal{H}_{\{\beta\}}=\mathcal{H}$. The partition function $\braket{\,}_{\left[\alpha,\beta\right]}$ is defined as an element of the tensor product $\mathcal{H}\otimes\mathcal{H}^{\,*}\otimes\mathrm{Func}\left(\mathrm{Metrics}_{\left[\alpha,\,\beta\right]}\right)$ satisfying the cutting axiom. This tensor product can be canonically identified with the space $\mathrm{End}\left(\mathcal{H}\right)\otimes\mathrm{Func}\left(\mathrm{Metrics}_{\left[\alpha,\,\beta\right]}\right)$. With this identification, the operation $\left(\cdot\right)_{\Gamma}$ becomes just a composition of linear operators. The moduli of the metrics $h$ on a segment $\left[\alpha,\beta\right]$ are just the lengths $l_{h}([\alpha,\beta])$ of the segment with respect to these metrics. Thus a partition function on the segment $\left[\alpha,\beta\right]$ is just such an endomorphism of the space $\mathcal{H}$ that for any point $\gamma\in\left[\alpha,\beta\right]$ the following relation holds: 
\begin{equation}
\label{eq:C.1.1}
\begin{aligned}
\braket{\,}_{\left[\alpha,\beta\right]_{h}}&=\braket{\,}_{\left[\alpha,\gamma\right]_{h_1}}\circ\braket{\,}_{\left[\gamma,\beta\right]_{h_2}}\,.
\end{aligned}
\end{equation}
Here $h_{1}=h|_{\left[\alpha,\,\gamma\right]}$ and $h_{2}=h|_{\left[\gamma,\,\beta\right]}$. From now on, we will omit the symbol $\circ$ that denotes the composition of linear operators. Equation~\eqref{eq:C.1.1} can be solved in the general case:
\begin{equation}
\label{eq:C.1.2}
\begin{aligned}
\braket{\,}_{\left[\alpha,\beta\right]_{h}}&=\mathrm{e}^{-l_{h}([\alpha,\beta]) H}\,,
\end{aligned}
\end{equation}
where $H$ is an arbitrary linear operator on the space $\mathcal{H}$. Self-conjugate operators $H$ correspond to unitary theories, but we can consider this construction in the general case. This operator is known as the Hamiltonian operator. The solution~\eqref{eq:C.1.2} is the well-known formula for the evolution operator in quantum mechanics.\footnote{In our convention, this is the imaginary time evolution operator, or the Euclidean evolution operator.} equation~\eqref{eq:C.1.1} is nothing more than the Dirac evolution \cite{DiracEvolution}. Thus, partition functions in one-dimensional FQFT are nothing more than Euclidean evolution operators.

We now turn to the study of local observables. According to the definition of local observable from section~\ref{subsec: Local Observables in Functorial QFT}, we must consider the segment $\left[\alpha,\beta\right]_{h}$ and then cut a one-dimensional disk $D_{r}^{(h)}(\tau)$ centered at $\tau\in\left(\alpha,\beta\right)$. The disk $D_{r}^{(h)}(\tau)$ centered at the point $\tau$ is a set of points at most $r$ away from the point $\tau$ in the sense of metric $h$. It is clear that such a disk is also a segment. Let the left edge of this disk be the point $\gamma_{1}(h,r,\tau)\in\left(\alpha,\beta\right)$ and the right edge be $\gamma_{2}(h,r,\tau)\in\left(\alpha,\beta\right)$. Note that after cutting the disk from the segment, it becomes a disjoint union of the two segments: $\left[\alpha,\beta\right]\backslash D_{r}^{(h)}(\tau)=\left[\alpha,\gamma_{1}\right]\sqcup\left[\gamma_{2},\beta\right]$. The appearance of two boundary components instead of one when the disk is cut is a feature of one-dimensional spacetime. It is useful to note here that $\gamma_{i}(h,r,\tau)$ is arranged so that when $r\rightarrow 0$ the boundary points $\gamma_{i}(h,r,\tau)\rightarrow \tau$. Now we consider the family of vectors $v_{r}$ in the boundary state space $\mathcal{H}_{\partial D_{r}^{(h)}(\tau)}$ such that there exists the limit
\begin{equation}
\label{eq:С.1.3}
\begin{aligned}
    \lim_{r\to 0}\left\{\left(\braket{\,}_{\left[\alpha,\beta\right]\backslash D_{r}^{(h)}(\tau)}\otimes v_{r}\right)_{\partial D_{r}^{(h)}(\tau)}\right\}\,.
\end{aligned}
\end{equation}
The product axiom states that
\begin{equation}
\label{eq:C.1.4}
\begin{aligned}
\braket{\,}_{\left[\alpha,\beta\right]\backslash D_{r}^{(h)}(\tau)}&=\mathrm{e}^{-l_{h}([\gamma_{2},\beta])H}\otimes\mathrm{e}^{-l_{h}([\alpha, \gamma_{1}])H}\in\mathrm{End}(\mathcal{H}\otimes\mathcal{H})\,.
\end{aligned}
\end{equation}
The space $\mathcal{H}_{\partial D_{r}^{(h)}(\tau)}$ can be canonically identified with $\mathrm{End}\left(\mathcal{H}\right)$. Given this identification, we will denote $v_{r}$ by $O_{r}$, the family of space endomorphisms $\mathcal{H}$. The limit~\eqref{eq:С.1.3} can then be expressed as follows:
\begin{equation}
\label{eq:C.1.5}
\begin{aligned}
    \lim_{r\to 0}\left\{\mathrm{e}^{-l_{h}([\gamma_{2},\beta])H}\,O_{r} \,\mathrm{e}^{-l_{h}([\alpha, \gamma_{1}])H}\right\}\,.
\end{aligned}
\end{equation}
One of the features of quantum mechanics is that the constant families $O_{r}=O$ are good.\footnote{We will see in the appendix~\ref{B.2} that this is no longer the case in two-dimensional field theory. There, families that do not depend on the radius of the cut disk are generally not good and do not define local observables.} Moreover, any observable in one-dimensional FQFT is implemented by a constant family. In fact, for a constant family, the limit~\eqref{eq:C.1.5} exists because there exist limits of $\gamma_{i}(h,r,\tau)$. By definition, the value of this limit is the correlator of the local observable defined by the constant family $O$:
\begin{equation}
\label{eq:C.1.6}
\begin{aligned}
\braket{O(\tau)}_{\left[\alpha,\,\beta\right]_{h}}&=\mathrm{e}^{-l_{h}([\tau,\beta])H}\,O \,\mathrm{e}^{-l_{h}([\alpha, \tau])H}\,.
\end{aligned}
\end{equation}
This is a well-known formula for the correlator of an observable in quantum mechanics.

In conclusion, we note a key feature of one-dimensional FQFT. The operator product of two observables in quantum mechanics is regular. In fact, let us consistently put local observables $O_{a}$ and $O_{b}$ at points $\tau_{1}$ and $\tau_{2}$ of the segment $\left[\alpha,\beta\right]_{h}$. Suppose $\tau_{1}>\tau_{2}$. The corresponding correlation function is given by the formula
\begin{equation}
\label{eq:C.1.7}
\begin{aligned}
\braket{O_{a}(\tau_{1})\,O_{b}(\tau_{2})}_{\left[\alpha,\,\beta\right]_{h}}&=\mathrm{e}^{-l_{h}([\tau_{1},\beta])H}\,O_{a} \,\mathrm{e}^{-l_{h}([\tau_{2},\tau_{1}])H}\,O_{b} \,\mathrm{e}^{-l_{h}([\alpha, \tau_{2}])H}\,.
\end{aligned}
\end{equation}
For a metric $h$ on the segment $\left[\alpha,\beta\right]$ with no singularities, the length $l_{h}([\tau_{2},\tau_{1}])$ tends to zero when $\tau_{2}\rightarrow \tau_{1}$. This is the reason why there are no ultraviolet divergences in the construction of the perturbative evolution operator in the standard approach to quantum mechanics.

\subsection{Perturbation Theory via Multiple Deformation in Quantum Mechanics}
\label{C.2}
Let us consider the theory on the segment $[\alpha,\beta]$ with the trivial metric, defined by the Hamiltonian $H\in\mathrm{End}\left(\mathcal{H}\right)$. Let us choose the canonical coordinate $\tau$ on this segment. Then the metric in this chart will have the form $\mathrm{d}\tau^{2}$, and the length of a segment is just the difference of the boundary coordinates of this segment:
\begin{equation}
\label{eq:C.2.1}
\begin{aligned}
l_{\mathrm{d}\tau^{2}}([\tau_{1},\tau_{2}])&=\tau_{2}-\tau_{1}\,.
\end{aligned}
\end{equation}
Thus the partition function on $[\tau_{1},\tau_{2}]\subset [\alpha,\beta]$ is given by the formula:
\begin{equation}
\label{eq:C.2.2}
\begin{aligned}
\braket{\,}_{[\tau_{1},\tau_{2}]}&=\mathrm{e}^{-(\tau_{2}-\tau_{1})H}\,.
\end{aligned}
\end{equation}
Let us now solve equation~\eqref{eq:C.1.1} in the neighborhood of the partition function~\eqref{eq:C.2.2}. Following the prescription from section~\ref{subsec: Deformation of the Partition Function} we can use any family of local observables to construct the deformation. Consider the family $\mathcal{O}=\left\{O_{a}\right\}_{a=1}^{n}$, then
\begin{equation}
\label{eq:C.2.3}
\begin{aligned}
\braket{\,}_{[\alpha,\beta]}^{(\mathrm{def})}(g,\mathcal{O})&=\braket{\,}_{[\alpha,\beta]}+g^{a}\cdot \int_{\alpha}^{\beta}\mathrm{d}\tau\,\braket{O_{a}(\tau)}_{[\alpha,\beta]}\,.
\end{aligned}
\end{equation}
This formula can be rewritten as follows:
\begin{equation}
\label{eq:C.2.4}
\begin{aligned}
\braket{\,}_{[\alpha,\beta]}^{(\mathrm{def})}(g,\mathcal{O})&=\mathrm{e}^{-(\beta-\alpha)H}+g^{a}\cdot \int_{\alpha}^{\beta}\mathrm{d}\tau\,\mathrm{e}^{-(\beta-\tau)H}\,O_{a}\,\mathrm{e}^{-(\tau-\alpha)H}\,.
\end{aligned}
\end{equation}
This partition function satisfies the cutting axiom by construction. This can be easily checked by choosing the point $\gamma\in(\alpha,\beta)$ and computing $\braket{\,}_{[\alpha, \gamma]}^{(\mathrm{def})}(g,\mathcal{O})\braket{\,}_{[\gamma,\beta]}^{(\mathrm{def})}(g,\mathcal{O})$ under the assumption that $g^{a}\cdot g^{b}=0$. Of course, this is a well-known formula from quantum mechanics, obtained by substituting the Hamiltonian $H+g^{a}\cdot O_{a}$ with the nilpotent $g^{a}$ into the exponential function.

Since observables do not need renormalization in quantum mechanics, we already know the observables space of the deformed theory: it is just the same one. This means that the family of observables of $\mathcal{O}$ of the theory $\braket{\,}_{[\alpha,\beta]}$ is also the family of observables of the theory $\braket{\,}_{[\alpha,\beta]}^{(\mathrm{def})}(g,\mathcal{O})$. Let us compute the correlator of the observable $O_{b}$ at point $\tilde{\tau}\in(\alpha,\beta)$ in the deformed theory. To do this, following the prescription from~\ref{subsec: Local Observables in Functorial QFT}, we should compute
\begin{equation}
\label{eq:C.2.5}
\begin{aligned}
\braket{O_{b}(\tilde{\tau})}_{[\alpha,\beta]}^{(\mathrm{def})}(g,\mathcal{O})=\braket{\,}_{[\tilde{\tau},\beta]}^{(\mathrm{def})}(g,\mathcal{O})\,O_{b}\braket{\,}_{[\alpha,\tilde{\tau}]}^{(\mathrm{def})}(g,\mathcal{O})=\braket{O_{b}(\tilde{\tau})}_{[\alpha,\beta]}\,+\\+\,g^{a}\cdot\int_{\tilde{\tau}}^{\beta}\mathrm{d}\tau\,\braket{O_{a}(\tau)O_{b}(\tilde{\tau})}_{[\alpha,\beta]}+g^{a}\cdot\int_{\alpha}^{\tilde{\tau}}\mathrm{d}\tau\,\braket{O_{b}(\tilde{\tau})O_{a}(\tau)}_{[\alpha,\beta]}\,.
\end{aligned}
\end{equation}
Here, it is convenient to introduce the time-ordering symbol $\mathcal{T}$. We define it as follows: 
\begin{equation}
\label{eq:C.2.6}
\begin{aligned}
\mathcal{T}\left\{\braket{O_{a}(\tau)O_{b}(\tilde{\tau})}_{[\alpha,\beta]}\right\}&=
\begin{cases}
   \braket{O_{a}(\tau)O_{b}(\tilde{\tau})}_{[\alpha,\beta]}\,,\tau>\tilde{\tau}\,;\\
   \braket{O_{b}(\tilde{\tau})O_{a}(\tau)}_{[\alpha,\beta]}\,,\tau<\tilde{\tau}\,.
 \end{cases}
\end{aligned}
\end{equation}
Note that this symbol is symmetric with respect to permutation $(\tau, a)\leftrightarrow (\tilde{\tau},b)$. Then the correlation function~\eqref{eq:C.2.5} can be rewritten as follows:
\begin{equation}
\label{eq:C.2.7}
\begin{aligned}
\braket{O_{b}(\tilde{\tau})}_{[\alpha,\beta]}^{(\mathrm{def})}(g,\mathcal{O})&=\braket{O_{b}(\tilde{\tau})}_{[\alpha,\beta]}+g^{a}\cdot \int_{\alpha}^{\beta}\mathrm{d}\tau\,\mathcal{T}\left\{\braket{O_{a}(\tau)O_{b}(\tilde{\tau})}_{[\alpha,\beta]}\right\}\,.
\end{aligned}
\end{equation}
Now using the correlator~\eqref{eq:C.2.7} we can deform the theory $\braket{\,}_{[\alpha,\beta]}^{(\mathrm{def})}(g,\mathcal{O})$ by the same family of local observables $\mathcal{O}$. Then we obtain
\begin{equation}
\label{eq:C.2.8}
\begin{aligned}
\braket{\,}_{[\alpha,\beta]}^{(2-\mathrm{def})}(g,\mathcal{O};\tilde{g},\mathcal{O})&=\braket{\,}_{[\alpha,\beta]}^{(\mathrm{def})}(g,\mathcal{O})+\tilde{g}^{a}\cdot \int_{\alpha}^{\beta}\mathrm{d}\tilde{\tau}\,\braket{O_{a}(\tilde{\tau})}_{[\alpha,\beta]}^{(\mathrm{def})}(g,\mathcal{O})\,.
\end{aligned}
\end{equation}
We can rewrite this partition function in terms of $\braket{\,}_{[\alpha,\beta]}$ and the correlators of $\mathcal{O}$:
\begin{equation}
\label{eq:C.2.9}
\begin{aligned}
\braket{\,}_{[\alpha,\beta]}^{(2-\mathrm{def})}(g,\mathcal{O};\tilde{g},\mathcal{O})=\braket{\,}_{[\alpha,\beta]}+(g^{a}+\tilde{g}^{a})\cdot \int_{\alpha}^{\beta}\mathrm{d}\tau\,\braket{O_{a}(\tau)}_{[\alpha,\beta]}\,+\\+\,\tilde{g}^{a}g^{b}\cdot\int_{\alpha}^{\beta}\mathrm{d}\tilde{\tau}\int_{\alpha}^{\beta}\mathrm{d}\tau\,\mathcal{T}\left\{\braket{O_{a}(\tau)O_{b}(\tilde{\tau})}_{[\alpha,\beta]}\right\}\,.
\end{aligned}
\end{equation}
Note that for a symmetric symbol $S_{ab}$ and first order nilpotents $g^{a},\tilde{g}^{a}$, it is true that
\begin{equation}
\label{eq:C.2.10}
\begin{aligned}
\tilde{g}^{a}g^{b}\cdot S_{ab}&=\frac{(\tilde{g}^{a}+g^{a})(\tilde{g}^{b}+g^{b})}{2}\cdot S_{ab}\,.
\end{aligned}
\end{equation}
Then we see that~\eqref{eq:C.2.9} depends only on the sum $g^{a}+\tilde{g}^{a}$, which we will denote by $g_{c}^{a}$. Rewriting the double-deformed partition function $\braket{\,}_{[\alpha,\beta]}^{(2-\mathrm{def})}$ in terms of $g_{c}^{a}$, we obtain
\begin{equation}
\label{eq:C.2.11}
\begin{aligned}
\braket{\,}_{[\alpha,\beta]}^{(2-\mathrm{def})}=\braket{\,}_{[\alpha,\beta]}+g_{c}^{a}\cdot \int_{\alpha}^{\beta}\mathrm{d}\tau\,\braket{O_{a}(\tau)}_{[\alpha,\beta]}\,+\\+\,\frac{g_{c}^{a}g_{c}^{b}}{2}\cdot\int_{\alpha}^{\beta}\mathrm{d}\tilde{\tau}\int_{\alpha}^{\beta}\mathrm{d}\tau\,\mathcal{T}\left\{\braket{O_{a}(\tau)O_{b}(\tilde{\tau})}_{[\alpha,\beta]}\right\}\,.
\end{aligned}
\end{equation}
This is a well-known formula for the second-order perturbative solution of the equation defining the evolution operator. 

Thus, we constructed the second-order perturbative solution of the cutting axiom as a double deformation of the exact solution. For a more coherent explanation of the connection between perturbation theory and multiple deformations, we refer to \cite{LosevGritskov}.
\section{Example II: Functorial Formulation of the Free Boson CFT}
\label{B}
\subsection{Partition Functions}
\label{B.1}
As an example of a functorial CFT, we implement the free boson theory in terms of a partition function satisfying Segal's axioms. We consider the free boson on a cylinder, on an annulus, and on a disk. 

The boundary state spaces are tensor products of $\mathcal{F}\otimes\bar{\mathcal{F}}\otimes \mathcal{H}_{0}$, where $\mathcal{F}$ and $\bar{\mathcal{F}}$ are Fock modules corresponding to chiral and anti-chiral modes, respectively, and $\mathcal{H}_{0}$ is the zero mode state space. The Fock module is organized in a standard way. There is an algebra of $\mathrm{U}(1)$-current generators $j_{n}, n\in\mathbb{Z}$:
\begin{equation}
\label{eq:B.1.1}
\begin{aligned}
\left[j_{m},j_{n}\right]&=m\cdot\delta_{m+n,0}\,.
\end{aligned}
\end{equation}
The Fock module has the highest weight vector $\ket{0}$, and it decomposes into a direct sum:
\begin{equation}
\label{eq:B.1.2}
\begin{aligned}
\mathcal{F}&=\bigoplus_{n\geq 0}\mathcal{F}_{n}\,,
\end{aligned}
\end{equation}
where $\mathcal{F}_{n}$ is the subspace of states with energy $n$. This is the state space of our theory if we neglect the zero modes. All basis states in $\mathcal{F}_{n}$, numbered by $\left\{\mu\right\}$, have the form
\begin{equation}
\label{eq:B.1.3}
\begin{aligned}
j_{\left\{\mu\right\}}\ket{0}&=j_{-\mu_{1}}\dots j_{-\mu_{k}}\ket{0},
\end{aligned}
\end{equation}
where $k\geq 0$, $\mu_{1}+\dots+\mu_{k}=n$, and $\mu_{1}\geq \mu_{2}\dots\geq \mu_{k}>0$. Generators of the conformal algebra are given by
\begin{equation}
\label{eq:B.1.4}
\begin{aligned}
L_{n}&=\frac{1}{2}\sum_{k\in\mathbb{Z}}:j_{-k}\,j_{k+n}:\,,\qquad \left[L_{m},L_{n}\right]=(m-n)\cdot L_{m+n}+\frac{m^{3}-m}{12}\cdot \delta_{m+n,0}\,,
\end{aligned}
\end{equation}
where we introduce the normal ordering of Heisenberg algebra generators, defined by the following way: $:j_{m}\,j_{n}:\,=j_{m}\,j_{n}$ if $m\leq n$ or $:j_{m}\,j_{n}:\,=j_{n}\,j_{m}$ if $m>n$. Thus, the spaces of boundary conditions are representations of the Virasoro algebra. The pairing of boundary spaces is done using the Shapovalov form. On the chiral sector, it is defined as follows:
\begin{equation}
\label{eq:B.1.5}
\begin{aligned}
\braket{0|0}&=1,\,\braket{\left\{\nu\right\}|\left\{\mu\right\}}=\bra{0}j_{\nu_{k}}...j_{\nu_{1}}\,j_{-\mu_{1}}...j_{-\mu_{m}}\ket{0}\,.
\end{aligned}
\end{equation}

Let us go to the first example. Consider a cylinder of length $H$ as the spacetime manifold $X$. The partition function is obviously given by the formula
\begin{equation}
\label{eq:B.1.6}
\begin{aligned}
\braket{\,}_{H}&=\mathrm{e}^{-H(L_{0}+\bar{L}_{0})}\,.
\end{aligned}
\end{equation}
It is clear that such a partition function satisfies the cutting axiom: if we cut a cylinder along the circle $\mathrm{S}^{1}$ so that we obtain two cylinders of lengths $H_{1}$ and $H_{2}$, then
\begin{equation}
\label{eq:B.1.7}
\begin{aligned}
\left(\braket{\,}_{H_{1}}\otimes \braket{\,}_{H_{2}}\right)_{\mathrm{S}^{1}}=\mathrm{e}^{-H_{1}(L_{0}+\bar{L}_{0})}\,\mathrm{e}^{-H_{2}(L_{0}+\bar{L}_{0})}=\mathrm{e}^{-(H_{1}+H_{2})(L_{0}+\bar{L}_{0})}=\underbrace{\mathrm{e}^{-H(L_{0}+\bar{L}_{0})}}_{\braket{\,}_{H}}\,.
\end{aligned}
\end{equation}

Our next example is the annulus. Since the annulus is a cylinder with exponential conformal factor of the metric, the partition function can be obtained directly from the previous example. However, one should be careful that the free boson has a non-zero central charge. Let the outer radius of the annulus be $R$ and the inner radius be $r$, then
\begin{equation}
\label{eq:B.1.8}
\begin{aligned}
\braket{\,}_{D_{R}\backslash D_{r}}&=\left(\frac{r}{R}\right)^{L_{0}+\bar{L}_{0}+\frac{1}{12}}\,.
\end{aligned}
\end{equation}
The appearance of $1/12$ in the exponent is due to the non-perturbative contribution to the conformal anomaly \cite{ConfAnomaly,ZZ} arising from the non-zero central charge. It is clear that this partition function satisfies the cutting axiom, for the same reasons as the partition function on the cylinder. For symmetry reasons, you can guess the partition function on a flat disk:
\begin{equation}
\label{eq:B.1.9}
\begin{aligned}
\braket{\,}_{D_{r}}&=r^{-\frac{1}{12}}\cdot\ket{0}\,.
\end{aligned}
\end{equation}
And then the cutting axiom is indeed satisfied:
\begin{equation}
\label{eq:B.1.10}
\begin{aligned}
\left(\braket{\,}_{D_{R}\backslash D_{r}}\otimes\braket{\,}_{D_{r}}\right)_{\partial D_{r}}&=\left(\frac{r}{R}\right)^{L_{0}+\bar{L}_{0}+\frac{1}{12}}\,r^{-\frac{1}{12}}\cdot\ket{0}=\underbrace{R^{-\frac{1}{12}}\cdot\ket{0}}_{\braket{\,}_{D_{R}}}\,.
\end{aligned}
\end{equation}

To simplify the formulas, we redefine $L_{0}\mapsto L_{0}-1/24$ and $\bar{L}_{0}\mapsto \bar{L}_{0}-1/24$. From the physical point of view, such a redefinition corresponds to the shift of the vacuum state energy. This is convenient because numerical factors such as $1/12$ in equations~\eqref{eq:B.1.8}-\eqref{eq:B.1.10} will disappear. Thus, the partition function $\braket{\,}_{D_{R}}$ on the disk will be equal to $\ket{0}$.
\subsection{Space of Local Observables and Current-Current OPE}
\label{B.2}
In this section we consider examples of local observables and operator product expansions in the free boson theory. We will consider the theory on the disk $D_{R}$. First, we will construct a holomorphic observables. As noted earlier, the space of boundary conditions is factorized such that $\mathcal{H}_{D_{R}}=\mathcal{F}\otimes\bar{\mathcal{F}}\otimes\mathcal{H}_{0}$. Therefore, the space of holomorphic good families for the partition function $\braket{\,}_{D_{R}\backslash D_{r}}$ has the form $r^{-L_{0}}\mathcal{F}$.

Let's consider the example of a good family, that defines holomorphic observable:
\begin{equation}
\label{eq:B.2.1}
\begin{aligned}
v_{r}&=r^{-1}\cdot j_{-1}\ket{0}=\braket{j\left(0\right)}_{D_{r}}\,.
\end{aligned}
\end{equation}
This local observable corresponds to a $\mathrm{U}(1)$-current placed in the center of the disk. The conformal dimension of this observable is $\Delta=1$. We can place this observable at some other point $z\in D_{R}$, obtaining the chiral one-point  correlator:
\begin{equation}
\label{eq:B.2.2}
\begin{aligned}
\braket{j\left(z\right)}_{D_{R}}&=\sum_{n\in \mathbb{Z}} z^{-n-1}\cdot R^{n}\cdot j_{n}\ket{0}=\sum_{n\geq1} z^{n-1}\cdot R^{-n}\cdot j_{-n}\ket{0}\,.
\end{aligned}
\end{equation}

Now we are ready to build the first example of the OPE. According to the procedure described in section~\ref{subsec: Operator Product Expansion} we should consider the following two-point correlation function:
\begin{equation}
\label{eq:B.2.3}
\begin{aligned}
\braket{j\left(z\right)j\left(0\right)}_{D_{R}}&=\left(\braket{j\left(z\right)}_{D_{R}\backslash D_{r}}\otimes \braket{j(0)}_{D_{r}}\right)_{\partial D_{r}}=\sum_{n\in \mathbb{Z}} z^{-n-1}\cdot R^{n-1}\cdot j_{n}\,j_{-1}\ket{0}\,.
\end{aligned}
\end{equation}
We can uncouple the summand for $n=1$
\begin{equation}
\label{eq:B.2.4}
\begin{aligned}
\braket{j\left(z\right)j\left(0\right)}_{D_{R}}&=z^{-2}\cdot\ket{0}+\sum_{n\geq1} z^{n-1}\cdot R^{-n-1}\cdot j_{-n}\,j_{-1}\ket{0}\,.
\end{aligned}
\end{equation}
Note that all but the first summand is regular at the point $z=0$. In addition, the sum in equation~\eqref{eq:B.2.4} is carried out using the correlation functions of some local observables of the free boson theory. Suppose $\braket{O_{a}(0)}_{D_{r}}$ is a chiral observable. Let us introduce its $\mathrm{U}(1)$ descendants by the following formula:
\begin{equation}
\label{eq:B.2.5}
\begin{aligned}
\braket{O_{a}^{\left\{\mu\right\}}(0)}_{D_{r}}&=r^{-L_{0}}\,j_{\left\{\mu\right\}}\,r^{L_{0}}\braket{O_{a}(0)}_{D_{r}}\,.
\end{aligned}
\end{equation}
Then we can rewrite~\eqref{eq:B.2.4} as
\begin{equation}
\label{eq:B.2.6}
\begin{aligned}
\braket{j\left(z\right)j\left(0\right)}_{D_{R}}&=z^{-2}\cdot\braket{1}_{D_{R}}+\sum_{n\geq0}z^{n}\cdot\braket{j^{\left\{n+1\right\}}(0)}_{D_{R}}\,,
\end{aligned}
\end{equation}
where $\braket{1}_{D_{R}}=\braket{\,}_{D_{R}}$ is the correlator of the identity observable that obviously coincides with the partition function. Thus, we have obtained an example of a decomposition~\eqref{eq:2.3.5}.

Similar observables and their OPE can be constructed for the antiholomorphic sector. By introducing antichiral modes, we can construct the following limit observable:
\begin{equation}
\label{eq:B.2.7}
\begin{aligned}
v_{r}&=r^{-2}\cdot j_{-1}\bar{j}_{-1}\ket{0}\,.
\end{aligned}
\end{equation}
Of course, the marginal sector of the free boson CFT is not exhausted by the local observable~\eqref{eq:B.2.7}. There are also the so-called vertex operators $V_{\alpha}$ \cite{MnevLect}. Their construction requires the use of the variable conjugate with zero mode $j_{0}$. The dimension of $V_{\alpha}$ is $\alpha^{2}/2$, so for $\alpha=\sqrt{2}$ they are marginals. Their algebra is organized as follows:
\begin{equation}
\label{eq:B.2.8}
\begin{aligned}
\braket{V_{\alpha}\left(z,\bar{z}\right)V_{\beta}\left(0\right)}_{D_{R}}&=|z|^{2\alpha\beta}\,\braket{V_{\alpha+\beta}\left(0\right)}_{D_{R}}+\,...
\end{aligned}
\end{equation}

\bibliographystyle{JHEP}
\bibliography{biblio}
\end{document}